\documentclass[smallcondensed,final]{svjour3}
\pdfoutput=1 

\usepackage{mathbbol}
\usepackage[numbers,sort&compress]{natbib}
\usepackage{dsfont,citesort}
\usepackage{dsfont}
\usepackage{color}
\usepackage{graphicx}
\usepackage{bm}
\usepackage{amsmath,amsfonts}
\usepackage{amssymb}
\usepackage{stmaryrd}
\usepackage{stackengine}

\newenvironment{remarks}[1][\textit{\textbf{Remarks:}}]{\begin{trivlist}
\item[\hskip \labelsep { #1}]}{\end{trivlist}}
\newenvironment{remarkk}[1][\textit{\textbf{Remark:}}]{\begin{trivlist}
\item[\hskip \labelsep { #1}]}{\end{trivlist}}
\newenvironment{fluctuation}[1][\textbf{Fluctuation}:]{\begin{trivlist}
\item[\hskip \labelsep { \textit{#1}}]}{\end{trivlist}}
\newenvironment{relaxation}[1][\textbf{Relaxation}:]{\begin{trivlist}
\item[\hskip \labelsep { \textit{#1}}]}{\end{trivlist}}

\newcommand{\sref}[1]{Section \ref{#1}}
\newcommand{\fref}[1]{Figure \ref{#1}}
\newcommand{\aref}[1]{Appendix \ref{#1}}

\newcommand{\rrangle}{\right \rangle}
\newcommand{\llangle}{\left \langle}
\newcommand{\rt}{\textrm{right}}
\newcommand{\lt}{\textrm{left}}

\journalname{J. Stat. Phys.}

\begin{document}

\title{Large deviations conditioned on large deviations II: Fluctuating hydrodynamics}

\author{Bernard Derrida   \and
        Tridib Sadhu
}

\institute{Bernard Derrida \at
              Coll\`{e}ge de France, 11 place Marcelin Berthelot, 75231 Paris Cedex 05 - France.\\
              bernard.derrida@college-de-france.fr
           \and
          Tridib Sadhu  \at
          Tata Institute of Fundamental Research, Homi Bhabha Road, Mumbai 400005.\\
          Coll\`{e}ge de France, 11 place Marcelin Berthelot, 75231 Paris Cedex 05 - France.\\
          tridib@theory.tifr.res.in
}

\maketitle

\begin{abstract}
For diffusive many-particle systems such as the SSEP (symmetric simple exclusion process) or independent particles coupled with reservoirs at the boundaries, we analyze the density fluctuations conditioned on current integrated over a large time. We determine the  conditioned large deviation function of density by a microscopic calculation. We then show that it can be expressed in terms of the solutions of Hamilton-Jacobi equations, which can be written for general diffusive systems using a fluctuating hydrodynamics description.

\end{abstract}

\section{Introduction}

In recent years, there has been a growing interest \cite{Jack2010,Orland2015,Hirschberg2015,Schutz2016,Popkov2011,Popkov2010,TOUCHETTE2009,Chetrite2015,Maes1999} in characterizing trajectories conditioned on rare events. Such questions appear in a wide range of topics, including protein folding \cite{Mey2014}, chemical reactions \cite{Delarue2017,Dykman1994}, stochastic models of gene expression \cite{Horowitz2017},
atmospheric activities \cite{Laurie2015}, glassy systems \cite{Garrahan2007,Garrahan2009}, disordered media \cite{Dorlas2001}, \textit{etc.}. One motivation is to find efficient algorithms where the rare events become typical such that they are computationally easy to generate \cite{Orland2015,Delarue2017,Limer2018,Giardina2011,cloning}. To calculate the probability of rare events one needs to understand how these rare events are created and how they relax. In these activities, a major interest concerns conditioning on an atypical value of an empirical observable, such as the integrated current, the empirical density, the entropy production, and the activity \cite{Lebowitz1999,TOUCHETTE2017,Maes2008,Maes20082,Mehl2008,Speck2012,Agranov2019}. How do the fluctuations get affected by such conditioning, and what is the effective dynamics in this conditioned ensemble? These are questions we address in this present work.

For Markov processes, in particular, for systems with few degrees of freedom and evolving according to a Langevin equation, it is known \cite{Sadhu20182,Jack2010,Jack2015,Chetrite2013,Chetrite2015,TOUCHETTE2017,Lecomte20072,Garrahan2007,Garrahan2009,Evans2004} how to describe the conditioned dynamics when the time window for the empirical observable is large. In this paper, we see how the same ideas \cite{Sadhu20182} can be extended to systems with  many degrees of freedom. 

\begin{figure}
 \centering \includegraphics[width=0.8\textwidth]{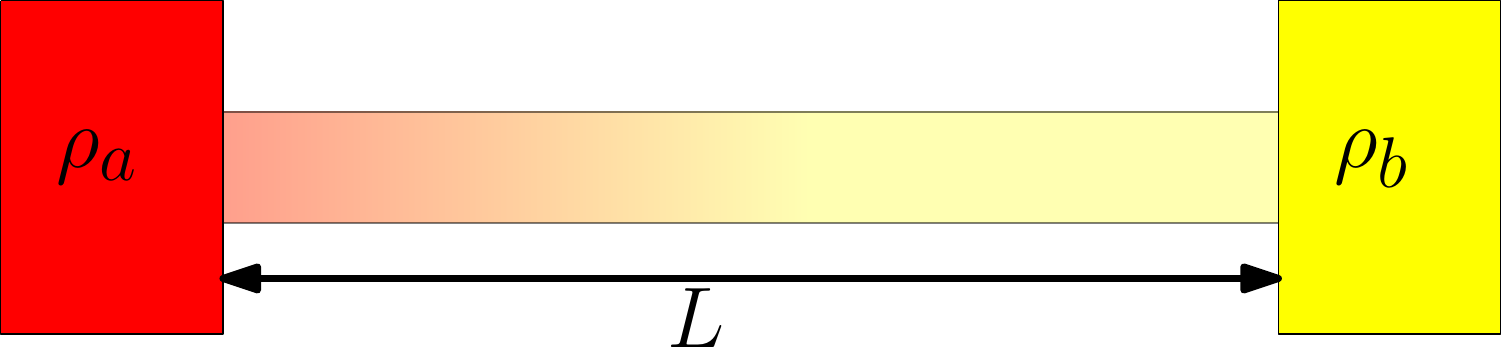}
  \caption{A diffusive system of length $L$ coupled with two reservoirs at different densities $\rho_a$ and $\rho_b$ at the boundary. \label{fig:fig1}}
\end{figure}

We work here in the steady state of a one-dimensional diffusive system of length $L$ coupled with reservoirs at the boundary, as shown in \fref{fig:fig1}. Under a diffusive re-scaling of space $i$ and time $\tau$, defining $\left(\frac{i}{L},\frac{\tau}{L^2}\right)\equiv (x,t)$ for large $L$, these systems are described \cite{Bertini2014,Bertini2009,Derrida2007,Eyink1990,Spohn1991,Sadhu2016} by a hydrodynamic density $\rho(x,t)$ and current $j(x,t)$ whose time evolution is governed by a fluctuating hydrodynamics equation
\begin{subequations}
\begin{equation}
\partial_t\rho(x,t)=-\partial_x j(x,t)\qquad \textrm{with}\qquad j(x,t)=-D(\rho(x,t))\; \partial_x\rho(x,t)+\eta(x,t)
\label{eq:fhd 1}
\end{equation}
where $D(\rho)$ is the diffusivity and $\eta(x,t)$ is a Gaussian white noise with zero mean and covariance 
\begin{equation}
\llangle \eta(x,t)\eta(x',t')\rrangle=\frac{1}{L}\sigma(\rho(x,t))\; \delta(x-x')\;\delta(t-t') \label{eq:covariance}
\end{equation}
where $\sigma(\rho)$ is the mobility. The density $\rho(x,t)$ could be the density of particles or of energy depending on the system. The microscopic details of the system are embedded in the functions $D(\rho)$ and $\sigma(\rho)$. The pre-factor $\frac{1}{L}$ in the covariance of $\eta(x,t)$ is due to coarsegraining, which makes the noise strength weak for large $L$ \cite{Sadhu2016,Bertini2014,Bertini2001,Derrida2007}.
\label{eq:fhd together}
\end{subequations}

Our interest is in the statistics of the steady state density $\rho(x)$ when conditioned on the integrated current $Q_T$ over a time interval $[0,T]$, for large $T$. Individual statistics of $\rho(x)$ and $Q_T$ have already been studied \cite{Derrida2007,Tailleur2008,Bertini2001,Bodineau2004,Bodineau2005,Derrida2001,Hurtado2010,Hurtado2014,Bertini2006Current,Bertini2007Current,Bertini2005Current}.
For the conserved dynamics \eqref{eq:fhd together}, when the density is bounded inside the bulk, the probability of $Q_T$ for large $T$ is independent of where the current is measured, and $P(Q_T)$ has a large deviation description \cite{Bodineau2004,Bodineau2005,Bertini2005Current}. On the other hand, in the steady state, the probability of $\rho(x)$ has also a large deviation form when the system size $L$ is large \cite{Derrida2001,Derrida2007,Tailleur2008,Bertini2001}. These two probabilities are given by
\begin{equation}
P(Q_T=q\, T)\sim e^{-T\,L\, \Phi(q)}\qquad \textrm{and} \qquad P[\rho(x)]\sim e^{-L\, \mathcal{F}[\rho(x)]}
\label{eq:ldf individual}
\end{equation}
for $T$ and $L$ both being large, where $\Phi(q)$ and $\mathcal{F}[\rho(x)]$ are the corresponding large deviation functions. 
Here, the precise meaning of the symbol $\sim$ is that $\lim_{L\rightarrow\infty}\lim_{T\rightarrow\infty}\frac{\log P(Q_T=qT)}{L\; T}=-\Phi(q)$ and 
$\lim_{L\rightarrow\infty}\frac{\log P[\rho(x)]}{L}=-\mathcal{F}[\rho(x)]$.

Their joint statistics, and equivalently, the conditioned probability of $\rho(x)$ for a given value of $Q_T$ depends on where the latter is measured. Taking this into account, we define an empirical observable
\begin{equation}
Q_T^{(\alpha)}=\int_0^1 dx\,\alpha(x) \int_0^T dt\, j(x,t)\qquad \textrm{with}\qquad \int_0^1 dx\, \alpha(x)=1.
\label{eq:QT}
\end{equation}
Here, $\alpha(x)$ is arbitrary except for the normalization\footnote{The normalization for $\alpha(x)$ in \eqref{eq:QT} ensures that the probability $P(Q_T^{(\alpha)}=qT)$ has the large deviation form \eqref{eq:ldf individual} with the same large deviation function $\Phi(q)$, independent of $\alpha(x)$.} in \eqref{eq:QT}.

We denote by $\mathcal{P}_t^{(\alpha)}[\rho(x)\vert Q]$ the conditioned probability of a density profile $\rho(x)$ at time $t$ given that $Q_T^{(\alpha)}$ takes value $Q$ while the system is in its steady state. 
For large $T$ and $L$, it has the large deviation form
\begin{align}
& \mathcal{P}_t^{(\alpha)}[\rho(x)\vert Q=qT]\sim  e^{-L \, \psi_t^{(\alpha)}[\rho(x), q]} \label{eq:cond prob micro}
\end{align}
where $\psi_t^{(\alpha)}[\rho(x), q]$ is the conditioned large deviation function.

\begin{figure}
 \centering \includegraphics[width=0.8\textwidth]{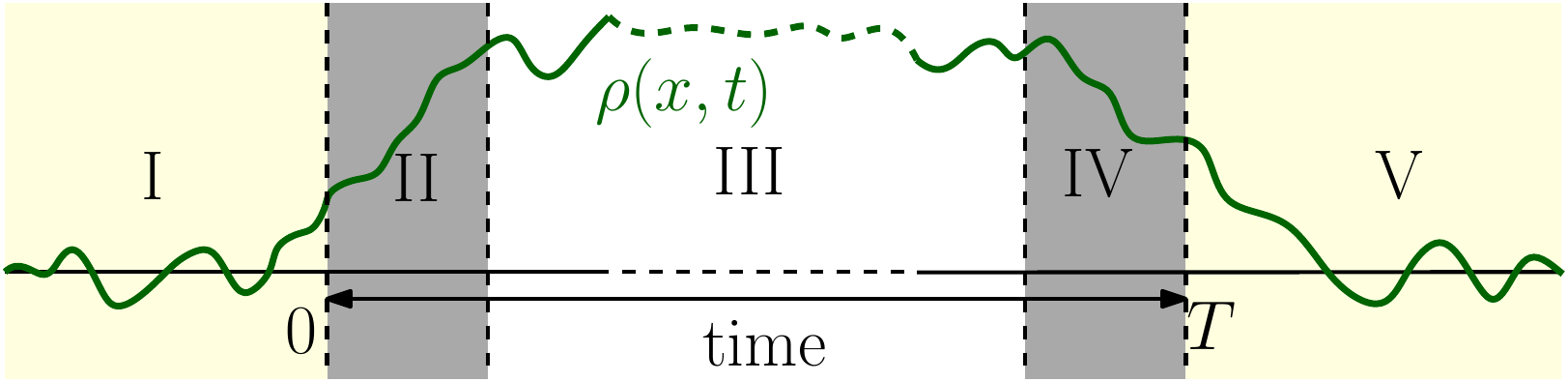}
  \caption{A schematic time evolution of the density profile $\rho(x,t)$ when conditioned on $Q_T^{(\alpha)}$ in \eqref{eq:QT} measured over a large time interval $[0,T]$. The system starts in its steady state far in the past, then reaches a quasi-stationary state for $1\ll t$ and $1\ll T-t$, and finally for $t\gg T$ it relaxes to its steady state. Different shaded regions denote different parts of the evolution: (I) $t<0$, (II) $t\ge 0$ but small, (III) $1\ll t$ and $1\ll T-t$, (IV) $T-t> 0$ but small, and (V) $t\ge T$. \label{fig:fig2}}
\end{figure}

Our goal is to determine this large deviation function $\psi_t^{(\alpha)}[\rho(x), q]$. In this paper, we will give this function at $t=0$, in the intermediate quasi-stationary state ($t\gg 1$ and $T-t\gg 1$), and at $t=T$ (indicated in \fref{fig:fig2}) for diffusing independent particles and for the symmetric simple exclusion process \cite{Derrida2007}. Our results for the latter are in a perturbation expansion in small density. A hydrodynamic description of these two examples is given by \eqref{eq:fhd together} with $D(\rho)=1$ and $\sigma(\rho)=2\rho$ for the diffusing independent particles, and $D(\rho)=1$ and $\sigma(\rho)=2\rho (1-\rho)$ for the symmetric simple exclusion process.

We shall start by a microscopic calculation on these models where we determine the conditioned probability in terms of the left and right eigenvectors associated to the largest eigenvalue of a tilted matrix \cite{Sadhu20182}. Then, taking a hydrodynamic limit of the conditioned probability we derive $\psi_t^{(\alpha)}[\rho(x), q]$. For example, in the quasi-stationary state (regime III in \fref{fig:fig2}), we will see that
\begin{equation}
\psi_t^{(\alpha)}[\rho(x), q]\equiv \psi_\textrm{qs}[\rho(x), q]=\psi_\lt^{(\alpha)}[\rho(x), q]+\psi_\rt^{(\alpha)}[\rho(x), q]
\label{eq:psi mid reln 1}
\end{equation}
up to an additive constant\footnote{We shall show that the $\alpha(x)$ dependence cancels in the expression \eqref{eq:psi mid reln 1} for $\psi_\textrm{qs}[\rho(x), q]$. In other regions of \fref{fig:fig2}, \textit{e.g.} at $t=0$ and $t=T$, $\psi_t^{(\alpha)}[\rho(x), q]$ depends on $\alpha(x)$.} (subscript ``qs'' denotes ``quasi-stationary''), where $\psi_\lt^{(\alpha)}$ and $\psi_\rt^{(\alpha)}$ are related \cite{Sadhu20182} to the left and right eigenvectors of a tilted matrix (see \sref{sec:cond ldf r l and p} for a precise definition). We will see later that these two functions $\psi_\lt^{(\alpha)}[\rho(x), q]$ and $\psi_\rt^{(\alpha)}[\rho(x), q]$ have in fact the following physical interpretation in terms of the large deviation function defined in \eqref{eq:cond prob micro}.
\begin{subequations}
\begin{align}
\psi_\lt^{(\alpha)}[\rho(x), q]=& \psi_0^{(\alpha)}[\rho(x), q]-\mathcal{F}[\rho(x)] \label{eq:psi left}\\
\psi_\rt^{(\alpha)}[\rho(x), q]=& \psi_T^{(\alpha)}[\rho(x), q]\label{eq:psi right}
\end{align}
up to an additive constant, where $\mathcal{F}[\rho(x)]$ is the unconditioned large deviation function of the density, as defined in \eqref{eq:ldf individual}.
\label{eq:psi right together}
\end{subequations}

\begin{figure}
 \centering \includegraphics[width=0.8\textwidth]{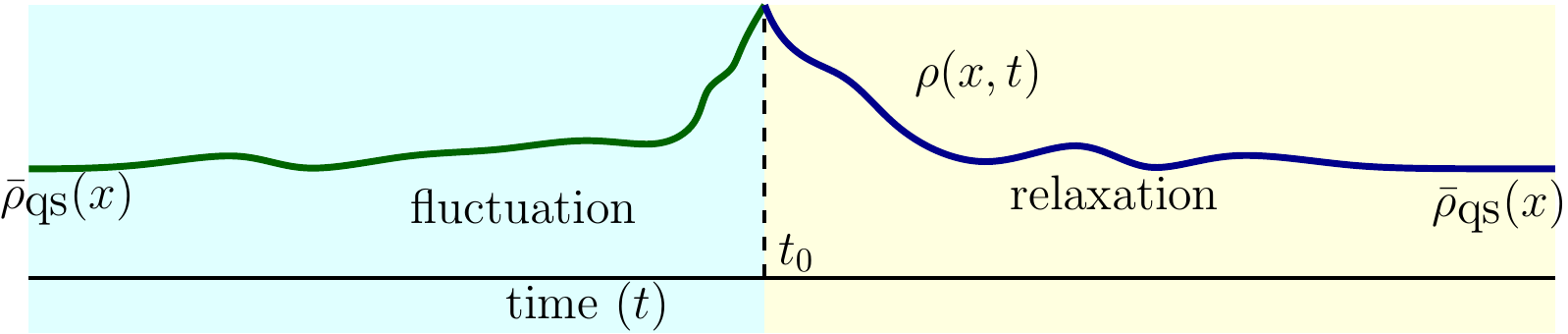}
  \caption{A schematic path $\rho(x,t)$ leading to a fluctuation in density at $t=t_0$ in the quasi-stationary regime, and its subsequent relaxation to the quasi-stationary density $\bar{\rho}_\textrm{qs}(x)$.  \label{fig:fig3}}
\end{figure}

In addition, we shall see that the conditioned dynamics for large $T$, can be effectively described by a fluctuating hydrodynamics equation with an additional driving field, which can be expressed in terms of $\psi_\lt^{(\alpha)}$ and $\psi_\rt^{(\alpha)}$. For example, in the quasi-stationary regime, the path $\rho(x,t)$ leading to a fluctuation at $t_0$ with $1\ll t_0$ and $1\ll T-t_0$ (illustrated in \fref{fig:fig3}), follows, for $t< t_0$,
\begin{subequations}
\begin{equation}
\partial_t\rho(x,t)=\partial_x \left\{D(\rho)\partial_x\rho-\sigma(\rho)\left(\Phi'(q)\, \alpha(x)+\partial_x \frac{\delta \psi^{(\alpha)}_\rt}{\delta \rho(x,t)}\right)+\eta(x,t)\right\}
\label{eq:J alpha expr fluc}
\end{equation}
with $\eta(x,t)$ being a Gaussian white noise of zero mean and covariance \eqref{eq:covariance}. Similarly, the path $\rho(x,t)$ for subsequent relaxation ($t\ge t_0$) follows
\begin{equation}
\partial_t\rho(x,t)=\partial_x \left\{D(\rho)\partial_x\rho-\sigma(\rho)\left(\Phi'(q)\, \alpha(x)-\partial_x \frac{\delta \psi^{(\alpha)}_\lt}{\delta \rho(x,t)} \right)+\eta(x,t)\right\}
\label{eq:J alpha expr relax}
\end{equation}
\end{subequations}

In the later part of this paper, we shall show that these results are consistent with a macroscopic approach starting from \eqref{eq:fhd together} for a rather general $D(\rho)$ and $\sigma(\rho)$. In this approach, $\psi_\lt^{(\alpha)}$ and $\psi_\rt^{(\alpha)}$ are solution of the Hamilton-Jacobi equations	
\begin{subequations}
\begin{align}
\int_0^1 dx\, \bigg\{ \frac{\sigma(\rho)}{2}  \left(\partial_x \frac{\delta \psi^{(\alpha)}_\lt}{\delta \rho(x,t)}-\Phi'(q)\alpha(x)+ \frac{D(\rho)\partial_x\rho}{\sigma(\rho)}\right)^2 &-\frac{(D(\rho)\partial_x\rho)^2}{2\sigma(\rho)}\bigg\}\cr & =\,\Phi'(q)\,q -\Phi(q) \label{eq:HJ1}\\
\int_0^1 dx\, \bigg\{ \frac{\sigma(\rho)}{2} \left(\partial_x \frac{\delta \psi^{(\alpha)}_\rt}{\delta \rho(x,t)}+\Phi'(q)\alpha(x)- \frac{D(\rho)\partial_x\rho}{\sigma(\rho)}\right)^2 &-\frac{(D(\rho)\partial_x\rho)^2}{2\sigma(\rho)}\bigg\}\cr & =\Phi'(q)\,q-\Phi(q) \label{eq:HJ2}
\end{align}
\end{subequations}
One should note that all our results for the conditioned process are in the large $T$ limit.

We present this paper in the following order. In \sref{sec:examples microscopic}, we discuss the microscopic framework for analyzing the conditioned probability in a general lattice gas model and introduce conditioned large deviation function in the hydrodynamic limit. Our calculation is in a weighted ensemble, which is known \cite{Sadhu20182,Jack2010,Jack2015,Chetrite2013,Chetrite2015,TOUCHETTE2017} to be equivalent to the conditioned process in the large $T$ limit (through an equivalence of ensembles). We use this procedure to derive $\psi_t^{(\alpha)}[\rho(x), q]$ for the diffusing independent particles in \sref{sec:ni micro analysis}, and for the symmetric simple exclusion process in \sref{sec:sep micro analysis}. Using this microscopic approach we describe the conditioned dynamics in \sref{sec:conditioned dynamics}. In \sref{sec:earlier results} and \sref{sec:examples macro}, we show how our expressions of the large deviation function $\psi_t^{(\alpha)}[\rho(x), q]$ fit with the Hamilton-Jacobi equations derived from a macroscopic approach starting from the fluctuating hydrodynamics description \eqref{eq:fhd together}.

\section{Microscopic analysis using the tilted matrix \label{sec:examples microscopic}}
\begin{figure}
 \centering \includegraphics[width=0.7\textwidth]{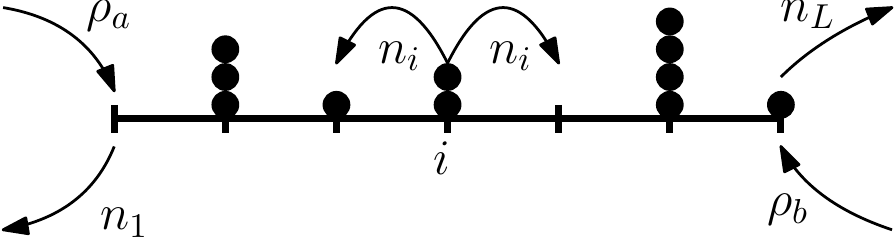}
\caption{Transition rates for independent particles on a one-dimensional chain coupled with reservoirs of density $\rho_a$ and $\rho_b$. The number of particles $n_i\ge 0$ at a site $i$ is arbitrary. \label{fig:fig6}}
\end{figure}

\begin{figure}
 \centering \includegraphics[width=0.7\textwidth]{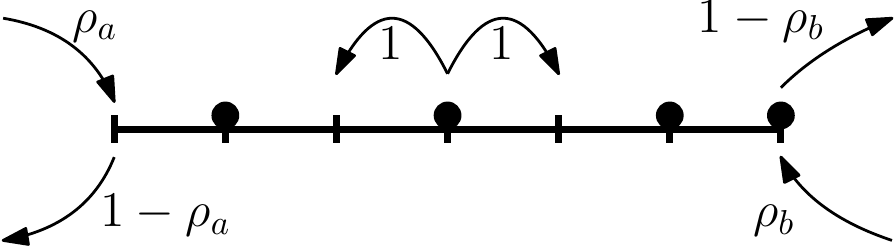}
  \caption{Transition rates of particles in the symmetric simple exclusion process on a one-dimensional chain coupled with reservoirs of density $\rho_a$ and $\rho_b$. A site is occupied at most by one particle at a time. \label{fig:fig7}}
\end{figure}

Let us first recall a few earlier results \cite{Sadhu20182,Jack2010,Jack2015,Chetrite2013,Chetrite2015,Chetrite20152,TOUCHETTE2017,Lecomte20072,Garrahan2009} in a Markov process conditioned on an empirical measure by writing them for the two examples: (a) diffusing independent particles and (b) the symmetric simple exclusion process. These are defined on a finite one-dimensional lattice of $L$ sites where particles jump between neighboring sites following a continuous time $\tau$ update rule (see \fref{fig:fig6} and \fref{fig:fig7}). The jump rates at the boundary correspond to coupling to reservoirs of density $\rho_a$ and $\rho_b$ \cite{Derrida2007}. In both examples, a configuration is specified by the set of occupation variables $\mathbf{n}\equiv\{n_1,\ldots,n_L\}$.

The microscopic analogue of the empirical observable \eqref{eq:QT} is
\begin{equation}
\mathcal{Q}_N^{(\boldsymbol{\lambda})}=\sum_{i=0}^{L}\lambda_i\,\times 
~\ensurestackMath{\stackanchor[1pt]{\text{number of jumps from site $i$ to $i+1$}}
  {\text{during the time interval $ [0,N]$ }}}
\label{eq:micro Q}
\end{equation}
where $\boldsymbol{\lambda}\equiv \{\lambda_0,\ldots,\lambda_L\}$ with $\lambda_i$ being real valued parameters. A jump from site $i+1$ to $i$ is counted as a jump from site $i$ to $i+1$ with a negative sign. (Here, $i=0$ denotes the left reservoir and $i=L+1$ denotes the right reservoir.) Similar to the condition in \eqref{eq:QT} we consider
\begin{equation}
\sum_{i=0}^{L}\lambda_i=1. \label{eq:sum alpha}
\end{equation}


\subsection{Tilted matrix}
Instead of conditioning on the value of $\mathcal{Q}_N^{(\boldsymbol{\lambda})}$, it is more convenient \cite{Sadhu20182,Jack2010,Jack2015,Chetrite2013,Chetrite2015,Chetrite20152} to work in the ensemble where events are weighted by a factor $e^{\kappa {\mathcal{Q}_N^{(\boldsymbol{\lambda})}}}$. By analogy \cite{Evans2004,Evans20042,Chetrite2013,Chetrite2015,TOUCHETTE2017,Jack2010,Sadhu20182} with equilibrium thermodynamics, we shall refer to this ensemble as the canonical ensemble and the ensemble where $\mathcal{Q}_N^{(\boldsymbol{\lambda})}$ is fixed as the micro-canonical ensemble. For large $N$, these two ensembles are equivalent (see \sref{sec:equivalence}).

In the canonical ensemble, we need to introduce the following tilted matrix \cite{Sadhu20182}
\begin{equation}
\mathcal{M}_\kappa^{(\boldsymbol{\lambda})}(\mathbf{n}',\mathbf{n})=\begin{cases}e^{\kappa\sum_{i=0}^{L}\lambda_i \, \mathcal{J}_{i}\left( \mathbf{n}',\mathbf{n}\right)}\mathcal{M}_0(\mathbf{n}',\mathbf{n}) \qquad & \textrm{for $\mathbf{n}'\ne \mathbf{n}$,}\cr
-\sum_{\mathbf{n}''\ne \mathbf{n}}\mathcal{M}_0(\mathbf{n}'',\mathbf{n}) \qquad &\textrm{for $\mathbf{n}'= \mathbf{n}$}\end{cases}
\label{eq:tilted M general}
\end{equation}
where $\mathcal{M}_0(\mathbf{n}',\mathbf{n})$ is the transition rate from configuration $\mathbf{n}$ to $\mathbf{n}'$ in the original (without condition) dynamics and for this transition,
\begin{equation*}
\mathcal{J}_{i}(\mathbf{n}',\mathbf{n})=\begin{cases}1, & \textrm{if a particle jumps from $i$ to $i+1$,}\cr -1, & \textrm{if a particle jumps from $i+1$ to $i$,}\cr 0, & \textrm{if no particle jumps between $i$ and $i+1$.} \end{cases}
\end{equation*}
The scaled cumulant generating function of $\mathcal{Q}_N^{(\boldsymbol{\lambda})}$, defined by
\begin{equation}
\mu(\kappa)=\lim_{N\rightarrow \infty}\frac{\log\left\langle e^{\kappa {\mathcal{Q}_N^{(\boldsymbol{\lambda})}}}\right\rangle}{N},
\label{eq:mu}
\end{equation}
is the largest eigenvalue of $\mathcal{M}_\kappa^{(\boldsymbol{\lambda})}$ \cite{Sadhu20182} such that
\begin{subequations}
\begin{align}
\sum_{\mathbf{n}}\mathcal{M}_\kappa^{(\boldsymbol{\lambda})}(\mathbf{n}',\mathbf{n})\,\mathcal{R}^{(\kappa,\boldsymbol{\lambda})}(\mathbf{n})=&\mu(\kappa)\,\mathcal{R}^{(\kappa,\boldsymbol{\lambda})}(\mathbf{n}')  \label{eq:eigenvalue equation right}\\
\sum_{\mathbf{n}'}\mathcal{L}^{(\kappa,\boldsymbol{\lambda})}(\mathbf{n}')\,\mathcal{M}_\kappa^{(\boldsymbol{\lambda})}(\mathbf{n}',\mathbf{n})=& \mu(\kappa)\,\mathcal{L}^{(\kappa,\boldsymbol{\lambda})}(\mathbf{n}) \label{eq:eigenvalue equation left}
\end{align}
where $\mathcal{R}^{(\kappa,\boldsymbol{\lambda})}$ and $\mathcal{L}^{(\kappa,\boldsymbol{\lambda})}$ are the associated right and left eigenvectors, respectively.\label{eq:eigenvalue equation together}
\end{subequations} 

At times of our interest, namely $\tau=0$, $\tau=N$, and in the quasi-stationary regime, the probability $P_\tau^{(\kappa,\boldsymbol{\lambda})}(\mathbf{n})$ of a configuration $\mathbf{n}$ in the canonical ensemble can be expressed (see \cite{Sadhu20182,Chetrite2015,Roche2004}) in terms of these eigenvectors up to normalization constants.
\begin{subequations}
\begin{itemize}
\item at time $\tau=0$,
\begin{equation}
P_0^{(\kappa,\boldsymbol{\lambda})}(\mathbf{n})=\mathcal{L}^{(\kappa,\boldsymbol{\lambda})}(\mathbf{n})\;\mathcal{R}^{(0,\boldsymbol{\lambda})}(\mathbf{n})
\label{eq:P0 micro}
\end{equation}
\item at time $\tau=N$,
\begin{equation}
P_N^{(\kappa,\boldsymbol{\lambda})}(\mathbf{n})=\mathcal{R}^{(\kappa,\boldsymbol{\lambda})}(\mathbf{n}) \label{eq:PT micro}
\end{equation}
\item and in the quasi-stationary regime, \textit{i.e.} $1\ll \tau$ with $1\ll N-\tau$,
\begin{equation}
P_\textrm{qs}^{(\kappa,\boldsymbol{\lambda})}(\mathbf{n})=\mathcal{L}^{(\kappa,\boldsymbol{\lambda})}(\mathbf{n})\,\mathcal{R}^{(\kappa,\boldsymbol{\lambda})}(\mathbf{n}). \label{eq:Pmid micro}
\end{equation}
\end{itemize}
\label{eq:PT micro togather can}
\end{subequations}

\subsection{Dependence on $\lambda$ \label{sec:symmetry}}

In our examples, the number of particles is conserved inside the bulk of the system, which means
\begin{equation*}
\mathcal{J}_{i-1}(\mathbf{n}',\mathbf{n})-\mathcal{J}_{i}(\mathbf{n}',\mathbf{n})=n_i'-n_i
\end{equation*}
for all $1\le i\le L$ and equivalently
\begin{equation*}
\mathcal{J}_{i}(\mathbf{n}',\mathbf{n})=\mathcal{J}_{0}(\mathbf{n}',\mathbf{n})-\sum_{j=1}^{i}(n_j'-n_j).
\end{equation*}
This conservation of particles leads to a symmetry of the tilted matrix and its eigenvectors. To see this, we use the above relation in \eqref{eq:tilted M general}. Here, using $\sum_{i=1}^{L}\lambda_i\sum_{j=1}^{i} n_j=\sum_{i=1}^{L} n_i\sum_{j\ge i}^{L}\lambda_j$ and the normalization \eqref{eq:sum alpha} we get 
\begin{align*}
\left[e^{\kappa \sum_{i=1}^{L}n_i' \sum_{j\ge i}^{L}\lambda_j }\right]\mathcal{M}_\kappa^{(\boldsymbol{\lambda})}(\mathbf{n}',\mathbf{n})\left[e^{-\kappa \sum_{i=1}^{L}n_i \sum_{j\ge i}^{L}\lambda_j }\right]=& \mathcal{M}_{\kappa}(\mathbf{n}',\mathbf{n})
\end{align*}
where $\mathcal{M}_{\kappa}$ is the tilted matrix for the case where $\lambda_0=1$ and $\lambda_i=0$ for rest of the sites. Denoting the eigenvectors of $\mathcal{M}_{\kappa}$ as $(\mathcal{R}^{(\kappa)},\,\mathcal{L}^{(\kappa)})$, we get
\begin{subequations}
\begin{align}
\mathcal{R}^{(\kappa,\boldsymbol {\lambda})}(\mathbf{n})\left[e^{\kappa\sum_{i=1}^{L}n_i\sum_{j\ge i}^{L}\lambda_j}\right] =& \mathcal{R}^{(\kappa)}(\mathbf{n}) \label{eq:symm2 R}\\
\mathcal{L}^{(\kappa,\boldsymbol {\lambda})}(\mathbf{n})\left[e^{-\kappa\sum_{i=1}^{L}n_i\sum_{j\ge i}^{L}\lambda_j}\right] =& \mathcal{L}^{(\kappa)}(\mathbf{n}).\label{eq:symm2 L}
\end{align}
Moreover, the eigenvalue $\mu(\kappa)$ is same in the two cases, which shows that it is independent of $\lambda$. \label{eq:symm together}
\end{subequations}

This gives the $\mathbf{\lambda}$-dependence of the probabilities \eqref{eq:PT micro togather can}. For example, using \eqref{eq:symm together} in \eqref{eq:Pmid micro} we see that in the quasi-stationary regime, the probability $P_\textrm{qs}^{(\kappa,\boldsymbol{\lambda})}$ is independent $\boldsymbol{\lambda}$ (given the normalization in \eqref{eq:sum alpha}),
\begin{equation}
P_\textrm{qs}^{(\kappa,\boldsymbol{\lambda})}(\mathbf{n})\equiv P_\textrm{qs}^{(\kappa)}(\mathbf{n}).
\label{eq:Pqs micro symm}
\end{equation}

\subsection{An equivalence of ensembles: canonical vs. micro-canonical \label{sec:equivalence}}

For large time $N$, there is \cite{Sadhu20182,Jack2010,Jack2015,Chetrite2013,Chetrite2015,TOUCHETTE2017} an equivalence between the canonical ensemble and the micro-canonical ensemble, where $\mathcal{Q}_N^{(\boldsymbol{\lambda})}$ is fixed. In this equivalence, the conditioned probability $\mathcal{P}_\tau^{(\boldsymbol{\lambda})}(\mathbf{n}\vert \mathcal{Q})$ of a configuration $\mathbf{n}$ at time $\tau$ given $\mathcal{Q}_N^{(\boldsymbol{\lambda})}=\mathcal{Q}$ is related to the probability $P_\tau^{(\kappa,\boldsymbol{\lambda})}(\mathbf{n})$ in the canonical ensemble by
\begin{equation}
\mathcal{P}_\tau^{(\boldsymbol{\lambda})}(\mathbf{n}\vert \mathcal{Q}=vN)\simeq P_\tau^{(\phi'(v),\boldsymbol{\lambda})}(\mathbf{n}) \qquad \textrm{for large time $N$,}
\label{equivalence arbitrary time macro}
\end{equation}
where $\phi(v)$ is the Legendre transform of the eigenvalue $\mu(\kappa)$, defined by
\begin{equation}
\phi(v)=v\, \kappa-\mu(\kappa)\qquad \textrm{with } \mu'(\kappa)=v.
\label{eq:equivalence micro}
\end{equation}
This also means \cite{Derrida2007,Bodineau2004,Bodineau2005,Bertini2005Current} that the probability of $\mathcal{Q}_N^{(\boldsymbol{\lambda})}$ has a large deviation form given by
\begin{equation}
P(\mathcal{Q}_N^{(\boldsymbol{\lambda})}=vN)\sim e^{-N\phi(v)} \qquad \textrm{ for large time $N$.}
\label{eq:P micro Q}
\end{equation}
In this work, our results for the micro-canonical ensemble are obtained from the canonical ensemble using the equivalence \eqref{equivalence arbitrary time macro}.

\subsection{Hydrodynamic limit \label{sec:hydrodynamic micro to macro}}
Our main interest is the probability in the hydrodynamic limit, which is defined in the scaled coordinates $(x,t)\equiv(\tfrac{i}{L},\tfrac{\tau}{L^2})$ for large $L$, with $N=L^2T$ and $\lambda_i=\frac{1}{L}\alpha(x)$ such that 
\begin{equation*}
\mathcal{Q}_N^{(\boldsymbol{\lambda})}\simeq L\,Q_T^{(\alpha)}\qquad \textrm{for large $L$,}
\label{eq:hydro limit first}
\end{equation*}
with $Q_T^{(\alpha)}$ given in \eqref{eq:QT}. In this hydrodynamic limit,
\begin{equation}
\phi\left(v=\frac{q}{L}\right)\simeq \frac{1}{L}\Phi(q),\qquad\qquad \mu(\kappa)\simeq \frac{1}{L}\chi(\kappa)
\label{eq:mu chi}
\end{equation}
and they are related (due to \eqref{eq:equivalence micro}) by
\begin{equation}
\Phi(q)=q\, \kappa-\chi(\kappa)\qquad \textrm{for } \chi'(\kappa)=q.
\label{eq:equivalence macro}
\end{equation}
This scaling \eqref{eq:mu chi} is well known \cite{Bodineau2004,Bodineau2005,Bertini2005Current,Bertini2007Current} and this will be confirmed in our examples.

For the probability of occupation variables $\mathbf{n}$, taking a hydrodynamic limit means \cite{Spohn1991,KIPNISLANDIM,Derrida2007} coarse-graining the system over boxes of width $w$ with $1\ll w\ll L$, such that each box has a uniform density $\rho(x)$ which varies smoothly on the hydrodynamic scale $x$. If $P(\mathbf{n})$ is the probability of a microscopic configuration $\mathbf{n}$, then the probability $P[\rho(x)]$ of a hydrodynamic density profile $\rho(x)$ is
\begin{equation}
\sum_{\mathbf{n}\in \rho(x)}P(\mathbf{n}) \simeq P[\rho(x)]
\label{eq:P hydro}
\end{equation}
where the summation is over all $\mathbf{n}$ that correspond to the profile $\rho(x)$ (see for example \sref{sec:ni micro analysis}).

The right eigenvector $\mathcal{R}^{(\kappa,\boldsymbol{\lambda})}$ has an interpretation of a probability (see \eqref{eq:PT micro}) and therefore, its hydrodynamic limit is similarly defined by
\begin{subequations}
\begin{align}
\sum_{\mathbf{n}\in \rho(x)}\mathcal{R}^{(\kappa,\boldsymbol{\lambda})}(\mathbf{n})\simeq r^{(\kappa,\alpha)}[\rho(x)]
\label{eq:R r}
\end{align}
In comparison, $\mathcal{L}^{(\kappa,\boldsymbol{\lambda})}$ by itself does not have an interpretation of a probability. Considering (\ref{eq:P0 micro},\,\ref{eq:Pmid micro}), we define the hydrodynamic limit for the left eigenvector as
\begin{align}
\mathcal{L}^{(\kappa,\boldsymbol{\lambda})}(\mathbf{n})\simeq \ell^{(\kappa,\alpha)}[\rho(x)]
\label{eq:L ell}
\end{align}
for each configuration $\mathbf{n}$ that corresponds to the profile $\rho(x)$. (Note that in contrast to \eqref{eq:R r} there is no summation over $\mathbf{n}$ in \eqref{eq:L ell}.)
\label{eq: R r L ell}
\end{subequations}

Using this construction of the hydrodynamic limit in \eqref{eq: R r L ell}, we see that the probability in \eqref{eq:PT micro togather can} leads to the probability of density $\rho(x)$, which 
\begin{itemize}
\item
at $t=0$ is given by
\begin{subequations}
\begin{equation}
P_{t=0}^{(\kappa,\alpha)}[\rho(x)]\simeq  \, \ell^{(\kappa,\alpha)}[\rho(x)]\;r^{(0,\alpha)}[\rho(x)],
\label{eq:P0 macro}
\end{equation}
\item
at $t=T$ is given by
\begin{equation}
P_{t=T}^{(\kappa,\alpha)}[\rho(x)]\simeq  \, r^{(\kappa,\alpha)}[\rho(x)], \label{eq:PT macro}
\end{equation}
\item
and in the quasi-stationary regime ($1\ll t$ and $1\ll T-t$) is given by
\begin{equation}
P_\textrm{qs}^{(\kappa,\alpha)}[\rho(x)]\equiv P_\textrm{qs}^{(\kappa)}[\rho(x)]\simeq \, \ell^{(\kappa,\alpha)}[\rho(x)]\; r^{(\kappa,\alpha)}[\rho(x)], \label{eq:Pmid macro}
\end{equation}
\label{eq:P0 macro together}
\end{subequations}
up to normalization constants.
\end{itemize}
\begin{remarkk}
Note that to obtain both \eqref{eq:P0 macro} and \eqref{eq:Pmid macro} we have replaced the sum $\sum_{\mathbf{n}\in \rho(x)}\mathcal{L}(\mathbf{n})\mathcal{R}(\mathbf{n})$ by $\ell[\rho(x)]\sum_{\mathbf{n}\in \rho(x)}\mathcal{R}(\mathbf{n})$ for large $L$. We will see that this relation is satisfied for the two models that we consider in this paper. It is expected to remain valid for more general diffusive systems \cite{Bodineau_private}.
\end{remarkk}

\subsection{Large deviation function \label{sec:cond ldf r l and p}}

In our examples we shall see that the hydrodynamic limit of the eigenvectors have a large deviation form given by
\begin{subequations}
\begin{align}
r^{(\kappa,\alpha)}[\rho(x)]\sim e^{-L\,\psi_\rt^{(\kappa,\alpha)}[\rho(x)]} \label{eq:R ldf}\\
\ell^{(\kappa,\alpha)}[\rho(x)]\sim e^{-L\,\psi_\lt^{(\kappa,\alpha)}[\rho(x)]} \label{eq:L ldf}
\end{align}\label{eq:R ldf together}
Using them in \eqref{eq:P0 macro together} gives\end{subequations}
\begin{equation}
\mathcal{P}_t^{(\kappa,\alpha)}[\rho(x)]\sim e^{-L\,\psi_t^{(\kappa,\alpha)}[\rho(x)]} \label{eq:psit general}
\end{equation}
with the large deviation function
\begin{subequations} 
\begin{itemize}
\item at $t=0$,
\begin{equation}
\psi_0^{(\kappa,\alpha)}[\rho(x)]=\psi_\lt^{(\kappa,\alpha)}[\rho(x)]+\psi_\rt^{(0,\alpha)}[\rho(x)]
\label{eq:psi 0 canonical}
\end{equation}
\item at $t=T$,
\begin{equation}
\psi_T^{(\kappa,\alpha)}[\rho(x)]=\psi_\rt^{(\kappa,\alpha)}[\rho(x)]
\label{eq:psi T canonical}
\end{equation}
\item and in the quasi-stationary regime, \textit{i.e.} $t\gg 1$ and $T-t\gg 1$, 
\begin{equation}
\psi_t^{(\kappa,\alpha)}[\rho(x)]\equiv\psi_{\textrm{qs}}^{(\kappa)}[\rho(x)]= \psi_\lt^{(\kappa,\alpha)}[\rho(x)]+\psi_\rt^{(\kappa,\alpha)}[\rho(x)]
\label{eq:psi mid reln 2}
\end{equation}
\end{itemize}
where all these equalities are up to an additive constant (independent of $\rho(x)$).
\label{eq:psi 0 canonical together}
\end{subequations}

\begin{remarkk}
From \eqref{eq:symm together} we see that the large deviation functions in \eqref{eq:R ldf together} have a simple dependence on $\alpha(x)$, given by
\begin{subequations}
\begin{align}
\psi^{(\kappa, \alpha)}_\rt[\rho(x)]=&V^{(\kappa)}_\rt[\rho(x)]+\kappa \int_0^1dx\, \rho(x)\int_x^1dy\,\alpha(y) \label{eq:psi right no alpha}\\
\psi^{(\kappa, \alpha)}_\lt[\rho(x)]=&V^{(\kappa)}_\lt[\rho(x)]-\kappa \int_0^1dx\, \rho(x)\int_x^1dy\,\alpha(y) \label{eq:psi left no alpha}
\end{align}
where $V^{(\kappa)}_\rt[\rho(x)]$ and $V^{(\kappa)}_\lt[\rho(x)]$ are the large deviation functions associated to $\mathcal{R}^{(\kappa)}$ and $\mathcal{L}^{(\kappa)}$ in \eqref{eq:symm together} (following a definition similar to \eqref{eq: R r L ell} and \eqref{eq:R ldf together}).
\label{eq:psi left no alpha together}
\end{subequations}
\end{remarkk}

\section{Independent particles \label{sec:ni micro analysis}}
In this section, we analyze the simple case of a system of independent particles with transition rates defined in \fref{fig:fig6}.
Considering the symmetry \eqref{eq:symm together} it is sufficient to analyze the case $\lambda_0=1$, and $\lambda_i=0$ for $i\ge 1$. In the rest of our analysis, we shall consider this case, unless explicitly stated otherwise.

In this case the tilted matrix \eqref{eq:tilted M general} is
\begin{align*}
\left[\mathcal{M}_\kappa \cdot \Omega\right](\mathbf{n})=&e^{\kappa }\rho_a \Omega(n_1-1,\ldots)+e^{-\kappa }(n_1+1)\Omega(n_1+1,\ldots)\cr
& +(n_L+1)\Omega(\ldots,n_L+1)+\rho_b \Omega(\ldots,n_L-1) \cr
 +\sum_{i=1}^{L-1}\bigg[(n_i+1)\Omega&(\ldots,n_i+1,n_{i+1}-1,\ldots)+(n_{i+1}+1)\Omega(\ldots,n_i-1,n_{i+1}+1,\ldots)\bigg]\cr
& -(\rho_a+2n_1+\ldots+2n_L+\rho_b)\Omega(\mathbf{n})
\end{align*}
where $\Omega(\mathbf{n})$ is the component of an arbitrary state vector $\boldsymbol{\Omega}$ in the configuration space.

The eigenvalue equations \eqref{eq:eigenvalue equation together} become
\begin{subequations}
\begin{align}
\left[\mathcal{M}_\kappa \cdot \mathcal{R}^{(\kappa)}\right](\mathbf{n})=&\mu(\kappa)\;\mathcal{R}^{(\kappa)}(\mathbf{n}) \label{eq:eigen eqn ni R}\\
\left[\mathcal{L}^{(\kappa)} \cdot \mathcal{M}_\kappa \right](\mathbf{n})=&\mu(\kappa)\;\mathcal{L}^{(\kappa)}(\mathbf{n}) \label{eq:eigen eqn ni L}
\end{align}
\label{eq:eigen eqn ni together}
\end{subequations}
One can check that the right and left eigenvectors are of the form
\begin{subequations}
\begin{align}
\mathcal{R}^{(\kappa)}(\mathbf{n})=&\prod_{i=1}^{L}\frac{a_i^{n_i}}{n_i!}e^{-a_i} \label{eq:R ni ansatz}\\
\mathcal{L}^{(\kappa)}(\mathbf{n})=&\prod_{i=1}^{L}b_i^{n_i} \label{eq:L ni ansatz}
\end{align}
\end{subequations}
where $(a_i, b_i)$ are positive numbers to be determined. With this ansatz the eigenvalue equations \eqref{eq:eigen eqn ni together} lead to
\begin{align*}
e^{-\kappa }a_1+& a_L-\rho_a-\rho_b-\mu(\kappa) +\sum_{i=1}^{L}\left[e^{\kappa \delta_{i,1} }\frac{a_{i-1}}{a_{i}}+\frac{a_{i+1}}{a_i}-2\right]n_i=0
\end{align*}
\begin{align*}
e^{\kappa}\rho_a b_1+& \rho_b b_L-\rho_a-\rho_b-\mu(\kappa) +\sum_{i=1}^{L}\left[e^{-\kappa \delta_{i,1} }\frac{b_{i-1}}{b_i}+\frac{b_{i+1}}{b_i}-2\right]n_i=0
\end{align*}
with $a_0=\rho_a$, $a_{L+1}=\rho_b$, and $b_0=1=b_{L+1}$. As the occupation variables $n_i$ are arbitrary, for the equations to be satisfied, their coefficients must vanish. This leads to a set of coupled linear equations for $a_i$ and $b_i$, which can be easily solved, and we get, for $1\le i\le L$,
\begin{subequations}
\begin{align}
a_i=& \rho_a\left(1-\frac{i}{L+1}\right)e^{\kappa}+\rho_b\frac{i}{L+1} \label{eq:ai ni}\\
b_i=& \left(1-\frac{i}{L+1}\right)e^{-\kappa}+\frac{i}{L+1} \label{eq:bi ni}
\end{align}
\end{subequations}
with the largest eigenvalue
\begin{equation}
\mu(\kappa)=\mu_\textrm{ni}(\kappa)=\frac{\rho_a}{L+1}\left(e^{\kappa}-1 \right)+\frac{\rho_b}{L+1}\left(e^{-\kappa}-1 \right)
\label{eq:mu ni}
\end{equation}

\subsection{Hydrodynamic limit}

In the hydrodynamic limit, from \eqref{eq:mu ni} it is easy to confirm \eqref{eq:mu chi}, which gives
\begin{equation}
\chi(\kappa)\equiv\chi_{\textrm{ni}}(\kappa)=\rho_a\left(e^{\kappa}-1 \right)+\rho_b\left(e^{-\kappa}-1 \right) \label{eq:chi ni explicit}
\end{equation}

For the hydrodynamic limit of the right eigenvector, defined in \eqref{eq:R r}, we decompose the system into $M=L/w$ boxes each containing $w$ sites and define
\begin{equation*}
r^{(\kappa)}(\rho_1,\ldots,\rho_{M})=\sum_{\mathbf{n}}\mathcal{R}^{(\kappa)}(\mathbf{n})
\end{equation*}
where the sum is over all configurations with $\rho_m w$ particles in the $m$-th box. Then, using \eqref{eq:R ni ansatz} we get
\begin{equation*}
r^{(\kappa)}(\rho_1,\ldots,\rho_{M})=\prod_{m=1}^{M}\left\{ \sum_{n_{m_1}}\cdots \sum_{n_{m_w}}\left[\frac{a_{m_1}^{n_{m_1}}}{n_{m_1}!}\cdots \frac{a_{m_w}^{n_{m_w}}}{n_{m_w}!}e^{-a_{m_1}\cdots-a_{m_w}}\right]\,\delta_{\rho_m w ,\sum_i n_{m_i}}\right\}
\end{equation*}
where $m_i$ denotes the site index of the $i$-th site of the $m$th box, and $\delta_{i,j}$ is the Kronecker delta. Using an identity
\begin{equation*}
\sum_{n_1=0}^{\infty}\sum_{n_2=0}^{\infty}\frac{a_{1}^{n_{1}}}{n_{1}!}\frac{a_{2}^{n_{2}}}{n_{2}!}\,\delta_{n,n_1+n_2}=\frac{(a_{1}+a_2)^{n}}{n!}
\end{equation*}
and that $a_{i}$ is slowly varying such that $a_{i}\simeq a(\frac{m w}{L})$ when the site $i$ is in the $m$-th box, and defining $\rho_{m}= \rho(\frac{m w}{L})$ we get
\begin{equation*}
r^{(\kappa)}(\rho_1,\ldots,\rho_{M})\simeq \prod_{m=1}^{M} \frac{\left[w\,a(\frac{m w}{L})\right]^{w\, \rho(\frac{m w}{L})}}{[w\, \rho(\frac{m w}{L})]!}\; e^{-w\,a(\frac{m w}{L})}
\end{equation*}
Then, for $1\ll w\ll L$, using the Stirling's formula we get (following the definition (\ref{eq:R r},\,\ref{eq:R ldf},\,\ref{eq:psi right no alpha}))
\begin{equation*}
r^{(\kappa)}(\rho_1,\ldots,\rho_{M})\simeq r^{(\kappa)}[\rho(x)]\sim e^{-L\,V_\rt^{(\kappa)}[\rho(x)]}
\end{equation*}
with 
\begin{subequations}
\begin{align}
V^{(\kappa)}_\rt[\rho(x)]= H_\textrm{ni}[\rho(x),a(x)] \label{eq:psi right ni}
\end{align}
where we defined
\begin{equation}
H_\textrm{ni}[\rho(x),a(x)]= \int_0^1 dx \left[\rho(x)\log\frac{\rho(x)}{a(x)} -\rho(x)+a(x)\right]
\end{equation}
and
\begin{align}
a(x)= \rho_a\left(1-x\right)e^{\kappa}+\rho_b x    \label{eq:a 0}
\end{align}
\label{eq:psi right ni together}
\end{subequations}

For the left eigenvector, the hydrodynamic limit \eqref{eq:L ell} is simple to construct from \eqref{eq:L ni ansatz}, which (following the definition (\ref{eq:L ell},\,\ref{eq:L ldf},\,\ref{eq:psi left no alpha})) leads to $\ell^{(\kappa)}[\rho(x)]\sim e^{-L\,V_\lt^{(\kappa)}[\rho(x)]}$ with
\begin{subequations}
\begin{align}
V^{(\kappa)}_\lt[\rho(x)]=\int_0^1 dx&\,\rho(x)\log\frac{1}{b(x)} 
\label{eq:psi left ni}
\end{align}
and
\begin{align}
b(x)= e^{-\kappa }(1-x)+ x   \label{eq:ell 0}
\end{align}
\label{eq:psi left ni together}
\end{subequations}

\begin{remarkk}
The $\kappa=0$ corresponds to the case without a condition on the empirical observable. So, the steady state large deviation function of density $\mathcal{F}_\textrm{ni}[\rho(x)]=V_\rt^{(0)}[\rho(x)]$ is
\begin{equation}
\mathcal{F}_\textrm{ni}[\rho(x)]=H_\textrm{ni}[\rho(x),\bar{\rho}_\text{free}(x)]\qquad \textrm{with }\quad\bar{\rho}_\text{free}(x)=\rho_a(1-x)+\rho_b x
\label{eq:ldf F}
\end{equation}
\end{remarkk}

\subsection{Large deviation function}
Expressions for $\psi^{(\kappa,\alpha)}_\lt$ and $\psi^{(\kappa,\alpha)}_\rt$ for an arbitrary $\alpha(x)$ follow from \eqref{eq:psi left no alpha together}. Using this result in \eqref{eq:psi 0 canonical together} we get the large deviation function at three different times, all of which are of the form
\begin{equation}
\psi_t^{(\kappa,\alpha)}[\rho(x)]=H_\textrm{ni}[\rho(x),\bar \rho_{t}^{(\kappa,\alpha)}(x)]
\end{equation}
with the average density profile
\begin{enumerate}
\item
at $t=0$,
\begin{subequations}
\begin{equation}
\bar \rho_{0}^{(\kappa,\alpha)}(x)=\bar{\rho}_\text{free}(x)\left[(1-x)e^{-\kappa\int_0^{x}dy\alpha(y)}+\, x\,e^{\kappa\int_x^{1}dy \alpha(y)}\right]
\end{equation}
\item at $t=T$,
\begin{equation}
\bar \rho_{T}^{(\kappa,\alpha)}(x)=\rho_a(1-x)e^{\kappa\int_0^{x}dy\alpha(y)}+\rho_b\, x\,e^{-\kappa\int_x^{1}dy \alpha(y)}
\label{eq:rho f sol}
\end{equation}
\item and in the quasi-stationary regime,
\begin{equation}
\bar \rho_\text{qs}(x)=\bar{\rho}_\text{free}(x)+x(1-x)\left[\rho_a\left(e^{\kappa}-1 \right)+\rho_b\left(e^{-\kappa}-1 \right)\right]
\label{eq:rhoth ni sol}
\end{equation}
\end{subequations}
\end{enumerate}

\section{Symmetric simple exclusion process \label{sec:sep micro analysis}}
In this section, we analyze the one-dimensional symmetric simple exclusion process with the transition rates defined in \fref{fig:fig7}. We indicate how to perform a low-density expansion. Our results will be limited to the first two terms in this expansion, although it is straightforward to extend our approach to higher orders. Considering the symmetry \eqref{eq:symm together} we will analyze only the case $\lambda_0=1$, and $\lambda_i=0$ for $i\ge 1$.

\subsection{A representation of the eigenvectors \label{sec:representation}}
A configuration can be specified by the position of the particles. In a configuration with $m$ occupied sites  $\{i_1,\ldots,i_m\}$ we denote the component of the eigenvectors as
\begin{equation}
\mathcal{R}^{(\kappa)}(\mathbf{n})\equiv \mathcal{R}^{(\kappa)}(i_1,\ldots,i_m)\qquad\textrm{and}\qquad  \mathcal{L}^{(\kappa)}(\mathbf{n})\equiv \mathcal{L}^{(\kappa)}(i_1,\ldots,i_m)
\label{eq:eigenvector sep representation}
\end{equation}
We normalize such that the component of both eigenvectors in the empty configuration is $1$. 

\subsection{A perturbation solution for small density \label{sec:perturbation expansion sep}}
For finite $L$, the Perron-Frobenius theorem \cite{VANKAMPEN2007193} assures that the largest eigenvalue of the tilted matrix is non-degenerate. Expressions of the eigenvalue and eigenvectors exist \cite{Lazarescu_2015,Vanicat}, but it is hard to extract from them the large scale behaviors. Here, we use a perturbation expansion in small $\rho_a$ and $\rho_b$, where it is possible to systematically solve the eigenvalue equation order by order. We write
\begin{subequations}
\begin{align}
\mu(\kappa)=&\mu_0(\kappa)+\mu_1(\kappa)+\mu_2(\kappa)+\cdots \label{eq:expansion mu sep}\\
\mathcal{R}^{(\kappa)}=& \mathcal{R}_0^{(\kappa)}+\mathcal{R}_1^{(\kappa)}+\mathcal{R}_2^{(\kappa)}+\cdots \label{eq:expansion r sep}\\
\mathcal{L}^{(\kappa)}=& \mathcal{L}_0^{(\kappa)}+\mathcal{L}_1^{(\kappa)}+\mathcal{L}_2^{(\kappa)}+\cdots \label{eq:expansion ell sep}
\end{align}
with increasing orders in $\rho_a$ and $\rho_b$.
\label{eq:expansion mu sep together}
\end{subequations} 

For $\rho_a=\rho_b=0$, all $\mathcal{R}_n^{(\kappa)}=0$ except for the empty configuration and one has $\mu_0(\kappa)=0$. It is also clear that at order $n$ in $\rho_a$ and $\rho_b$, the $\mathcal{R}_n^{(\kappa)}$ of configurations with more than $n$ occupied sites vanish. Therefore,
\begin{align*}
&\mathcal{R}_0^{(\kappa)}(i)=0,\cr
&\mathcal{R}_0^{(\kappa)}(i,j)=\mathcal{R}_1^{(\kappa)}(i,j)=0,\cr
&\mathcal{R}_0^{(\kappa)}(i,j,k)=\mathcal{R}_1^{(\kappa)}(i,j,k)=\mathcal{R}_2^{(\kappa)}(i,j,k)=0,
\end{align*}
and so on. Here we present the solution up to only the second order.

The equations one needs to solve up to the second order in $\rho_a$ and $\rho_b$ are given in the \aref{sec:eigen equation sep}. Since at this order $\mathcal{R}_2^{(\kappa)}$ of configurations with $3$ or more occupied sites vanish, the hierarchy closes. For arbitrary $L$, we get the solution
\begin{subequations}
\begin{align}
\mu_1(\kappa)=&\frac{\rho_a}{L+1}\left(e^{\kappa}-1 \right)+\frac{\rho_b}{L+1}\left(e^{-\kappa}-1 \right) \label{eq:mu mico 1}\\
\mu_2(\kappa)=&- \frac{\left(e^\kappa-1\right)^2}{6(L+1)^2}\left[2L\left(\rho_a^2+\rho_a\rho_be^{-\kappa}+\rho_b^2e^{-2\kappa}\right) + \rho_a^2+4\rho_a\rho_be^{-\kappa}+\rho_b^2e^{-2\kappa}\right] \label{eq:mu mico 2}
\end{align}
in agreement with an earlier result \cite{Roche}. The right eigenvector, up to the second order, is given by \label{eq:mu mico together}
\end{subequations}
\begin{subequations}
\begin{align}
\mathcal{R}_1^{(\kappa)}(i)=& \,e^{\kappa}\rho_a\left(1-\frac{i}{L+1}\right)+\rho_b\frac{i}{L+1} \label{eq:r mico 1}\\
\mathcal{R}_2^{(\kappa)}(i)=& \,e^{\kappa}\rho_a^2\left(1-\frac{i}{L+1}\right)+\rho_b^2\frac{i}{L+1} \cr
&-\left(e^\kappa\rho_a-\rho_b\right)^2\frac{i}{6L}\left(1-\frac{i}{L+1}\right)\left[e^{-\kappa}\left(2-\frac{i}{L+1}\right) +1+\frac{i}{L+1}\right]\\
\mathcal{R}_2^{(\kappa)}(i,j) &-  \mathcal{R}_1^{(\kappa)}(i)\mathcal{R}_1^{(\kappa)}(j)= -\frac{\left(\rho_a e^\kappa-\rho_b\right)^2}{(L+1)}\frac{i}{L}\left(1-\frac{j}{L+1}\right) \qquad \textrm{for $j\ge i$}\label{eq:rij mico 1}
\end{align}
\label{eq:R sep together}
\end{subequations}
As the probabilities \eqref{eq:PT micro togather can} are always a product of a left and a right eigenvector, we will only need the following orders for the left eigenvector.
\begin{subequations}
\begin{align}
\mathcal{L}_0^{(\kappa)}(i)=& e^{-\kappa}\left(1-\frac{i}{L+1}\right)+\frac{i}{L+1} \label{eq:l mico 1}\\
\mathcal{L}_1^{(\kappa)}(i)=& \left(e^{-\kappa}-1\right)^2\frac{i}{6L}\left(\frac{i}{L+1}-1\right)  \left[e^{\kappa}\rho_a\left(2-\frac{i}{L+1}\right)+\rho_b\left(1+\frac{i}{L+1}\right)\right]\\
\mathcal{L}_0^{(\kappa)}(i,j)&- \mathcal{L}_0^{(\kappa)}(i)\; \mathcal{L}_0^{(\kappa)}(j)= -\frac{\left( e^{-\kappa}-1\right)^2}{(L+1)}\frac{i}{L}\left(1-\frac{j}{L+1}\right) \label{eq:lij mico 1} \qquad \textrm{for $j\ge i$}
\end{align}
\label{eq:L sep together}
\end{subequations}
\begin{remarks}~\\
\begin{itemize}
\item
For the symmetric simple exclusion process, the eigenvalue equation can be systematically solved to arbitrary order. The ``miracle'' which makes an explicit solution possible in practice is that at every order, the eigenvectors are low degree polynomials of the site indices $i$. This is special to the exclusion process, and may not apply to other diffusive systems.
\item
At the first order in $\rho_a$ and $\rho_b$, the expressions of eigenvalue and eigenvectors coincide with that of the independent particles case in \sref{sec:ni micro analysis}. 
\end{itemize}
\end{remarks}

\subsection{Cumulants of the occupation variable \label{sec:cumulants}}
One can then derive the cumulants of the occupation variables from the probability \eqref{eq:PT micro togather can}. For example, at time $\tau=N$, using  \eqref{eq:PT micro} and the representation \eqref{eq:eigenvector sep representation} we get
\begin{align*}
\llangle n_i \rrangle=&\frac{1}{\mathcal{N}}\bigg(\mathcal{R}^{(\kappa)}(i)+\sum_{j\ne i}\mathcal{R}^{(\kappa)}(i,j)+\cdots\bigg)\cr
\llangle n_i n_j \rrangle=&\frac{1}{\mathcal{N}}\bigg(\mathcal{R}^{(\kappa)}(i,j)+\sum_{k\ne i,j}\mathcal{R}^{(\kappa)}(i,j,k)+\cdots\bigg)
\end{align*}
where the normalization $\mathcal{N}=1+\sum_i\mathcal{R}^{(\kappa)}(i)+\sum_i\sum_{j\ne i}\mathcal{R}^{(\kappa)}(i,j)+\cdots$.

Using this with the perturbation solution of the eigenvectors we can construct a perturbation expansion of the cumulants for the low density limit. For example, at the second order in $\rho_a$ and $\rho_b$, the average occupation of a site $i$ at time $\tau=N$ is
\begin{subequations}
\begin{equation}
\llangle n_i \rrangle=\mathcal{R}_1^{(\kappa)}(i)+\mathcal{R}_2^{(\kappa)}(i)-\left(\mathcal{R}_1^{(\kappa)}(i)\right)^2+\sum_{j\ne i}\left(\mathcal{R}_2^{(\kappa)}(i,j)-  \mathcal{R}_1^{(\kappa)}(i)\mathcal{R}_1^{(\kappa)}(j)\right)+\cdots
\label{eq:cumulant1 T micro}
\end{equation}
and the connected correlation 
\begin{equation}
\llangle n_i n_j\rrangle_c=\mathcal{R}_2^{(\kappa)}(i,j)-  \mathcal{R}_1^{(\kappa)}(i)\mathcal{R}_1^{(\kappa)}(j)+\cdots
\label{eq:cumulant2 T micro}
\end{equation}
For time $\tau=0$ and for the quasi-stationary state (see \eqref{eq:PT micro togather can}) we can similarly write the average occupations and their correlations by replacing $\mathcal{R}^{(\kappa)}$ by $\mathcal{R}^{(0)}\mathcal{L}^{(\kappa)}$ or $\mathcal{R}^{(\kappa)}\mathcal{L}^{(\kappa)}$ in \eqref{eq:cumulant1 T micro together}.
\label{eq:cumulant1 T micro together}
\end{subequations}

\subsection{Hydrodynamic limit \label{sec:hydro for sep}}
It is straightforward to take the hydrodynamic limit of the expressions \eqref{eq:R sep together} and \eqref{eq:L sep together}, which can be used to derive the hydrodynamic limit of the cumulants of the occupation variables. For large $L$, we define
\begin{align}
\ell^{(\kappa)}(x)& \simeq \mathcal{L}_0^{(\kappa)}(x\,L)+\mathcal{L}_1^{(\kappa)}(x\,L)\cr
&=e^{-\kappa}(1-x)+ x-\frac{1}{6}(e^{-\kappa}-1)^2x(1-x)\left[e^{\kappa}\rho_a(2-x)+\rho_b(1+x)\right] \label{eq:ell sep expression}
\end{align}
and
\begin{align}
g_\lt^{(\kappa)}(x,y) & \simeq L\;\left[\mathcal{L}_0^{(\kappa)}(xL,yL)- \mathcal{L}_0^{(\kappa)}(xL)\; \mathcal{L}_0^{(\kappa)}(yL)\right]\cr
&=-(e^{-\kappa}-1)^2 x(1-y)\qquad \text{for $y\ge x$.} \label{eq:gleft sep expression}
\end{align}
Similarly, from \eqref{eq:R sep together} we define
\begin{align}
r^{(\kappa)}(x)& \simeq \mathcal{R}_1^{(\kappa)}(x\,L)+\mathcal{R}_2^{(\kappa)}(x\,L)\cr
&=e^{\kappa}\left(\rho_a+\rho_a^2\right)(1-x)+\left(\rho_b+\rho_b^2 \right) x \cr
&\qquad \qquad -\frac{1}{6}(e^{\kappa}\rho_a-\rho_b )^2 x (1-x)\left[1+x+\left(2-x\right)e^{-\kappa} \right]
\end{align}
and
\begin{align}
g_\rt^{(\kappa)}(x,y) & \simeq L\;\left[\mathcal{R}_2^{(\kappa)}(xL,yL)- \mathcal{R}_1^{(\kappa)}(xL)\; \mathcal{R}_1^{(\kappa)}(yL)\right]\cr
&=-(e^{\kappa}\rho_a-\rho_b )^2 x(1-y)\qquad \text{for $y\ge x$.} \label{eq:gright sep}
\end{align}
\begin{remarkk}
In taking the hydrodynamic limit of the perturbation expansion we have assumed that the limits of large $L$ and small density can be exchanged. We will see that the resulting large deviation functions are consistent with the macroscopic analysis in \sref{sec:sep macro}.
\end{remarkk}

\subsubsection{Cumulants of density \label{sec:cumulants hydro sep}}
At time $\tau=0$, in the quasi-stationary regime, and at $\tau=L^2 T$ (the hydrodynamic time $t=T$) the cumulants of the occupation variable are of the form (see \eqref{eq:cumulant1 T micro together})
\begin{subequations}
\begin{align}
\langle n_{xL}\rangle &\simeq \bar\rho(x)=u(x)\left[1-u(x)\right]+\int_0^1dy\;c(x,y)\\
\langle n_{xL} \; n_{yL}\rangle_c & \simeq \frac{1}{L}c(x,y)  
\end{align}
\label{eq:rho c}
\end{subequations}
for large $L$, up to the second order in $\rho_a$ and $\rho_b$, where
\begin{itemize}
\item at $t=0$,
\begin{subequations}
\begin{align}
u(x)=&\ell^{(\kappa)}(x)r^{(0)}(x)\\
c(x,y)=& \ell^{(\kappa)}(x)\ell^{(\kappa)}(y)g_\rt^{(0)}(x,y)+r^{(0)}(x)r^{(0)}(y)g_\lt^{(\kappa)}(x,y)
\end{align}
\end{subequations}
\item in the quasi-stationary state,
\begin{subequations}
\begin{align}
u(x)=&\ell^{(\kappa)}(x)r^{(\kappa)}(x)\\
c(x,y)=& \ell^{(\kappa)}(x)\ell^{(\kappa)}(y)g_\rt^{(\kappa)}(x,y)+r^{(\kappa)}(x)r^{(\kappa)}(y)g_\lt^{(\kappa)}(x,y)
\end{align}
\label{eq:cumulants qs}
\end{subequations}
\item at $t=T$,
\begin{align}
u(x)=r^{(\kappa)}(x)\qquad \text{and}\qquad c(x,y)= g_\rt^{(\kappa)}(x,y) \label{eq:cumulants T}
\end{align}
\end{itemize}
Explicit expressions of the cumulants are given in \aref{app:cumulants}.

\subsubsection{Large deviation function \label{sec:psi moment reln}}
Here, we show how to derive a small density expansion of the large deviation function in \eqref{eq:psit general} from the above expansion of the cumulants of the density.

Take an arbitrary probability distribution $P(\mathbf{n})$ where the occupation variables are either $0$ or $1$. Then, one can write
\begin{equation*}
\prod_{i=1}^{L} e^{h_i n_i}= \ \prod_{i=1}^{L} \left[1 + n_i \left(e^{h_i}-1\right) \right]
\end{equation*}
where $h_i$ is a real valued parameter. If one expands this product, averages over $P(\mathbf{n})$, and uses the fact that cumulants of order $k$ scale like $\rho_a$ and $\rho_b$ to the power $k$, then one gets, for finite $L$,
\begin{align*}
\log \left\langle \prod_{i=1}^{L} e^{h_i n_i}\right\rangle =& \sum_{i=1}^{L}\langle n_i \rangle \left(e^{h_i} -1 \right)
- {1 \over 2} 
\sum_{i=1}^{L}\langle n_i \rangle^2 \left(e^{h_i} -1 \right)^2 \cr
& + \sum_{i=1}^{L}\sum_{j>i}^{L} \llangle n_i n_j \rrangle_c
\left(e^{h_i} -1 \right)
\left(e^{h_j} -1 \right) + \cdots
\end{align*}

In the hydrodynamic limit, when $h_{i} \simeq h(\frac{i}{L})$, 
$\llangle n_{i} \rrangle \simeq \bar{\rho}(\frac{i}{L})$, and $\llangle n_{i} n_{j} \rrangle_c \simeq {1 \over L} c(\frac{i}{L},\frac{j}{L})$, we get
for the generating functional
\begin{align*}
\log \left\langle \prod_{i=1}^{L} e^{h_i n_i}\right\rangle \simeq L \, \mathcal{G}[h(x)]
\end{align*}
with
\begin{align}
\mathcal{G}[h(x)] = & 
\int_0^1 dx \,  \left( e^{h(x)} -1\right) \bar{\rho}(x)- {1 \over 2} \int_0^1 dx \,  \left( e^{h(x)} -1\right)^2 \bar{\rho}(x)^2 \cr & +  \int_0^1 dx \int_x^1 dy 
\left( e^{h(x)} -1\right)\left( e^{h(y)} -1\right)  c(x,y)+ \cdots
\label{eq:G cum gen fnc}
\end{align}

Then, $P[\rho(x)]$ has \cite{Derrida2007,TOUCHETTE2009} the large deviation form \eqref{eq:psit general} with the large deviation function $\psi[\rho(x)]$ given by the Legendre transformation
\begin{equation}
\psi[\rho(x)] = \int_0^1 dx \, h(x) \, \rho(x) \,  - \,  \mathcal{G}[h(x)]
\label{eq:FG 1}
\end{equation}
where $h(x)$ is the solution of $\frac{\delta \mathcal{G}[h]}{\delta h(x)}=\rho(x)$.

The small density expansion of (\ref{eq:G cum gen fnc},\,\ref{eq:FG 1}) is then straightforward to get from the perturbation expansion of the cumulants. It gives
\begin{subequations}
\begin{equation}
\psi[\rho(x)]\simeq H_\textrm{sep}[\rho(x),\bar \rho(x), c(x,y) ] \label{eq:F sep series}
\end{equation}
where up to the second order in $\rho_a$ and $\rho_b$,
\begin{align}
H_\textrm{sep}[\rho(x),\bar \rho(x), c(x,y) ]=&\int_0^1 dx \left[ \rho(x)\log \frac{\rho(x)}{\bar{\rho}(x)}+(1-\rho(x))\log \frac{1-\rho(x)}{1-\bar{\rho}(x)}\right]\cr & -\frac{1}{2}\int_0^1dx \int_0^1dy \left( \frac{\rho(x)}{\bar{\rho}(x)}-1\right) \left( \frac{\rho(y)}{\bar{\rho}(y)}-1\right)c(x,y) \label{eq:H sep}
\end{align}
The expression \eqref{eq:F sep series together} remains valid at all times with $\bar{\rho}(x)$ and $c(x,y)$ being replaced by their expressions in \sref{sec:cumulants hydro sep}. For the unconditioned case, where $\bar{\rho}(x)$ and $c(x,y)$ are given by \eqref{eq:rho c} for $\kappa=0$, one can verify that this perturbation expansion result is consistent with previously known expressions \cite{Derrida2007,Bertini2001,Derrida2001}.
\label{eq:F sep series together}
\end{subequations}

\subsubsection{Small density expansion of $\chi(\kappa)$, $V_\lt^{(\kappa)}$ and $V_\rt^{(\kappa)}$}
It is straightforward to take the hydrodynamic limit of \eqref{eq:mu mico together}, which confirms \eqref{eq:mu chi} at the level of a small density expansion with
\begin{equation}
\chi(\kappa)=\rho_a\left(e^{\kappa}-1 \right)+\rho_b\left(e^{-\kappa}-1 \right)-\frac{1}{3}\left(1-e^{-\kappa} \right)^2\left(e^{2\kappa}\rho_a^2+e^\kappa \rho_a\rho_b+\rho_b^2 \right)+\cdots \label{eq:chi sep expansion}
\end{equation}
This is in agreement with earlier findings in \cite{Roche2004,Bodineau2004}.

For $\lambda_0=1$ and $\lambda_i=0$ for $i\ge 1$, the large deviation functions $\psi_\lt^{(\kappa,\alpha)}\equiv V_\lt^{(\kappa)}$ and $\psi_\rt^{(\kappa,\alpha)}\equiv V_\rt^{(\kappa)}$ (see \eqref{eq:psi left no alpha together}). An expression for $V_\lt^{(\kappa)}$ can be derived by taking the hydrodynamic limit (\ref{eq:L ell},\,\ref{eq:L ldf}) of $\mathcal{L}^{(\kappa)}$. 
For a configuration $\mathbf{n}$, where the occupation variables are either $0$ or $1$, we write
\begin{equation*}
\log\mathcal{L}^{(\kappa)}=\sum_i n_i\; \log\mathcal{L}^{(\kappa)}(i)+\sum_{i<j}n_in_j \;\log\left[1+\frac{\mathcal{L}^{(\kappa)}(i,j)-\mathcal{L}^{(\kappa)}(i)\mathcal{L}^{(\kappa)}(j)}{\mathcal{L}^{(\kappa)}(i) \mathcal{L}^{(\kappa)}(j)}\right]+\cdots \label{eq:psi definition micro to macro}
\end{equation*}
using the representation in \eqref{eq:eigenvector sep representation}.
Then, from the perturbation expansion (\ref{eq:expansion ell sep},\,\ref{eq:L sep together}) and taking the hydrodynamic limit (\ref{eq:L ell},\,\ref{eq:L ldf}) we get
\begin{equation}
V_\lt^{(\kappa)}[\rho]=\int_0^1 dx \; \rho(x)\log \frac{1}{\ell^{(\kappa)}(x)}-\frac{1}{2}\int_0^1dx \int_0^1dy \, \frac{\rho(x)\; \rho(y)}{\ell^{(\kappa)}(x)\;\ell^{(\kappa)}(y)}\;g^{(\kappa)}_{\lt}(x,y)+\cdots  \label{eq:psi left compact formula}
\end{equation}
where $\ell^{(\kappa)}(x)$ and $g_\lt^{(\kappa)}(x,y)$ are defined in \eqref{eq:ell sep expression} and \eqref{eq:gleft sep expression}.

In comparison, it is harder to derive an expression for $V_\rt^{(\kappa)}$ by taking the hydrodynamic limit of the expression \eqref{eq:R sep together} for $\mathcal{R}^{(\kappa)}$. It is much easier to derive using the relation \eqref{eq:psi T canonical} and the result (\ref{eq:rho c},\,\ref{eq:F sep series together}) at time $T$. This gives
\begin{subequations}
\begin{align}
V_\rt^{(\kappa)}[\rho(x)]&=H_\textrm{sep}\left[\rho(x),\bar{\rho}_T(x), g_\rt^{(\kappa)}(x,y) \right]\\
\text{with}\quad\bar{\rho}_T(x)&=r^{(\kappa)}(x)\left(1-r^{(\kappa)}(x)\right) + \int_0^1dy\,g_\rt^{(\kappa)}(x,y)
\end{align}
up to the second order in $\rho_a$ and $\rho_b$, and an additive constant. 
\label{eq:psif compact}
\end{subequations}

\begin{remarkk}
We have checked (details in \aref{sec:mlf}) that the result \eqref{eq:psi left compact formula}, and the expression of $\psi_\textrm{qs}^{(\kappa)}$ and $\psi_T^{(\kappa)}$ obtained from \eqref{eq:F sep series together} satisfy the relation \eqref{eq:psi mid reln 2}.
\end{remarkk}

\section{Effective dynamics \label{sec:conditioned dynamics}}

In the canonical ensemble, the biased dynamics is Markovian and one can write \cite{Sadhu20182,Chetrite2015,TOUCHETTE2017} the transition rates in terms of the tilted Matrix. For example, in the quasi-stationary regime, the transition rate $W_{\textrm{qs}}^{(\kappa,\boldsymbol{\lambda})}(\mathbf{n}',\mathbf{n})$ from a microscopic configuration $\mathbf{n}$ to another configuration $\mathbf{n}'$ is given by \cite{Sadhu20182}
\begin{subequations}
\begin{equation}
W_{\textrm{qs}}^{(\kappa,\boldsymbol{\lambda})}(\mathbf{n}',\mathbf{n})=
\frac{\mathcal{L}^{(\kappa,\boldsymbol{\lambda})}(\mathbf{n}')}{\mathcal{L}^{(\kappa,\boldsymbol{\lambda})}(\mathbf{n})}\,\mathcal{M}_\kappa^{(\boldsymbol{\lambda})}(\mathbf{n}',\mathbf{n}) \qquad \textrm{for $\mathbf{n}'\ne \mathbf{n}$.}
\label{eq:dynamics qs}
\end{equation}
This means, a spontaneous fluctuation at time $\tau_0$ in the quasi-stationary regime ($1\ll \tau_0$ and $1 \ll N-\tau_0$), relaxes following this dynamics \eqref{eq:dynamics qs}. 

Similarly, the path leading to a spontaneous fluctuation can be described by a time reversal of \eqref{eq:dynamics qs}. The transition rate $\mathbb{W}_{\textrm{qs}}^{(\kappa,\boldsymbol{\lambda})}(\mathbf{n}',\mathbf{n})$ of this time-reversed process can be constructed \cite{Stroock14} using the quasi-stationary distribution \eqref{eq:Pmid micro}, which gives
\begin{equation}
\mathbb{W}_{\textrm{qs}}^{(\kappa,\boldsymbol{\lambda})}(\mathbf{n}',\mathbf{n})=
\frac{\mathcal{R}^{(\kappa,\boldsymbol{\lambda})}(\mathbf{n}')}{\mathcal{R}^{(\kappa,\boldsymbol{\lambda})}(\mathbf{n})}\,\mathcal{M}_\kappa^{(\boldsymbol{\lambda})}(\mathbf{n},\mathbf{n}') \qquad  \textrm{for $\mathbf{n}'\ne \mathbf{n}$.}
\label{eq:time reverse dynamics qs}
\end{equation}
\end{subequations}

For the two examples considered in this paper (the independent particles and the symmetric simple exclusion process), it is straightforward to see that the effective dynamics \eqref{eq:dynamics qs} with \eqref{eq:tilted M general} correspond to re-weighting the jump rates of particles (see \fref{fig:fig6} and \fref{fig:fig7}): the jump rate for a particle from site $i$ to $i+1$ is weighted by a factor $e^{E_i}$, whereas the jump rate from $i+1$ to $i$ is weighted by $e^{-E_i}$, where
\begin{subequations}
\begin{equation}
E_i(\mathbf{n})=\kappa \, \lambda_i+\log \frac{\mathcal{L}^{(\kappa,\boldsymbol{\lambda})}(\widehat{\mathbf{n}})}{\mathcal{L}^{(\kappa,\boldsymbol{\lambda})}(\mathbf{n})}\qquad \qquad \textrm{for all $0\le i\le L$,}
\label{eq:E}
\end{equation}
with
\begin{equation}
\widehat{n}_j=n_j-\delta_{j,i}+\delta_{j,i+1} \qquad\qquad  \textrm{for all $1\le j\le L$.}
\label{eq:n hat n}
\end{equation}
Similar re-weighting of jump rates can be seen for the time reversed dynamics \eqref{eq:time reverse dynamics qs}.
We note that, in general, the jump rates for the dynamics are non-local functions of the occupation variables $\mathbf{n}$.
\label{eq:E together}
\end{subequations}

\subsubsection*{Hydrodynamic limit}
In the large $L$ limit, when $x=\frac{i}{L}$ and $\lambda_i=\frac{1}{L}\alpha(x)$, \eqref{eq:E together} becomes
\begin{align}
E_i(\mathbf{n})\simeq \frac{1}{L}e(x)\qquad &\textrm{with}\qquad e(x)=\kappa\,\alpha(x)-\partial_x\frac{\delta \psi_\lt^{(\kappa,\alpha)}}{\delta \rho(x)}.\label{eq:small e}
\end{align}

In a hydrodynamic description \eqref{eq:fhd together}, the effect of such a weak bias can be incorporated (see \cite{Bodineau2010} for another example) within the linear response theory, where $e(x)$ acts as an external driving field. This leads to the following dynamics.
\begin{subequations}
\begin{relaxation}\label{proposition 1}
A spontaneous fluctuation of hydrodynamic density in the quasi-stationary state relaxes following $\partial_t\rho(x,t)=-\partial_x j(x,t)$ with
\begin{equation}
j(x,t)=-D(\rho(x,t))\partial_x\rho(x,t)+\sigma(\rho(x,t))\left\{\kappa\,\alpha(x)-\partial_x\frac{\delta \psi_\lt^{(\kappa,\alpha)}}{\delta \rho(x,t)} \right\}+\eta(x,t)
\label{eq:fhd relax}
\end{equation}
where $\eta(x,t)$ is a Gaussian white noise of zero mean and covariance \eqref{eq:covariance}.
\end{relaxation}

A somewhat similar analysis based on \eqref{eq:time reverse dynamics qs} leads to the following dynamics:
\begin{fluctuation}\label{proposition 2}
The path leading to a fluctuation in the quasi-stationary state is described by $\partial_t \rho(x,t)=-\partial_x j(x,t)$ with
\begin{equation}
j(x,t)=-D(\rho(x,t))\partial_x\rho(x,t)+\sigma(\rho(x,t))\left\{\kappa\,\alpha(x)+\partial_x\frac{\delta \psi_\rt^{(\kappa,\alpha)}}{\delta \rho(x,t)} \right\}+\eta(x,t)
\label{eq:fhd fluc}
\end{equation}
\end{fluctuation}
\label{eq:fhd fluc together}
\end{subequations}
 
\begin{remarkk}
The time evolution of the most probable density profile leading to a fluctuation and its subsequent relaxation are the zero noise case of \eqref{eq:fhd fluc} and \eqref{eq:fhd relax}. We have verified this explicitly for the independent particles starting from their microscopic dynamics. For the symmetric simple exclusion process, we checked this up to the second order in a low density expansion.
\end{remarkk}

\section{Macroscopic analysis \label{sec:earlier results}}
The two examples discussed in \sref{sec:ni micro analysis} to \sref{sec:conditioned dynamics} are governed \cite{Bertini2002,Bodineau2004,Derrida2007,Sadhu2016}, for large $L$, by the fluctuating hydrodynamics equation \eqref{eq:fhd together}. Our goal here is to show that the large deviation functions (\ref{eq:psi right ni together},\,\ref{eq:psi left ni together}) and (\ref{eq:psi left compact formula},\,\ref{eq:psif compact}) are consistent with a macroscopic approach. Besides this, the macroscopic analysis applies for a general class of models where the microscopic details enter in the terms $D(\rho)$ and $\sigma(\rho)$.

Much of the results can be infered by drawing an analogy of \eqref{eq:fhd together} to a Langevin equation in the weak noise limit, as in \cite{Sadhu20182}. In this analogy, a simple quantity is the generating function, which for \eqref{eq:fhd together} is defined by
\begin{equation}
G_T^{(\kappa,\alpha)}[r(x), s(x)]=\int dQ\, e^{L\kappa \, Q}P^{(\alpha)}_T[r(x),Q\vert s(x)]
\label{eq:G def}
\end{equation}
where $P^{(\alpha)}_T[r(x),Q\vert s(x)]$ is the joint probability of a density profile $\rho(x,T)=r(x)$ at the hydrodynamic time $t=T$, and  $Q_T^{(\alpha)}$ in \eqref{eq:QT} to take value $Q$ given the density $\rho(x,0)=s(x)$ at $t=0$.
Similarly to the Langevin equation \cite{Sadhu20182}, one expects, for large $L$ and $T$, the generating function to have the form
\begin{equation}
G_T^{(\kappa,\alpha)}[r(x), s(x)]\sim e^{T\, L\, \chi(\kappa)-L\, \psi^{(\kappa, \alpha)}_\rt[r(x)]-L\, \psi^{(\kappa, \alpha)}_\lt[s(x)]}
\label{eq:G ldf}
\end{equation}
where $\chi(\kappa)$, $\psi^{(\kappa, \alpha)}_\rt$, and $\psi^{(\kappa, \alpha)}_\lt$ are the same
quantities as in \eqref{eq:mu chi} and \eqref{eq:R ldf together}. Starting from \eqref{eq:G ldf} we now obtain a variational formulation  as in \cite{Bertini2014,Bertini2009,Bertini2001,Derrida2007}. 

\subsection{A variational formulation \label{sec:variational approach}}
For large $L$, the probability of a certain time evolution of $\rho(x,t)$ and $j(x,t)$ inside the time window $[0,T]$, which follows \eqref{eq:fhd together} is given by \cite{Derrida2007,Bertini2014}
\begin{equation}
P[\rho(x,t),j(x,t)]\sim \exp\left[ -L\int_0^1 dx \int_0^T dt \frac{(j(x,t)+D(\rho)\partial_x\rho(x,t))^2}{2\sigma(\rho)} \right]
\label{eq:prob weight}
\end{equation}
where $\sim$ means that sub-leading terms in large $L$ are neglected.
Using this, the generating function \eqref{eq:G def} can be written as a path-integral
\begin{subequations}
\begin{equation}
G_T^{(\kappa,\alpha)}[r(x), s(x)]\sim \int \mathcal{D}[\rho,j]e^{L S_T^{(\kappa,\alpha)}[\rho,j]}
\label{eq:path integral}
\end{equation}
with
\begin{equation}
S_{t_f-t_i}^{(\kappa,\alpha)}[\rho,j]=\int_{t_i}^{t_f}dt\int_0^1dx\left\{\kappa \,\alpha(x) j(x,t)- \frac{(j(x,t)+D(\rho)\partial_x\rho(x,t))^2}{2\sigma(\rho)}\right\}
\label{eq:action}
\end{equation}\label{eq:action together}
\end{subequations}
with $t_i=0$ and $t_f=T$.
The path integral in \eqref{eq:path integral} is over all paths $\{\rho(x,t),j(x,t)\}$ satisfying $\partial_t\rho=-\partial_x j$ with the initial density profile $\rho(x,0)=s(x)$ and the final density profile $\rho(x,T)=r(x)$. 

For large $L$ and $T$, \textit{assuming a single optimal path}, we get the large deviation form \eqref{eq:G ldf} with
\begin{equation}
T\, \, \chi(\kappa)-\, \psi^{(\kappa, \alpha)}_\rt[r(x)]-\, \psi^{(\kappa, \alpha)}_\lt[s(x)]=\max_{\rho,j} S_T^{(\kappa,\alpha)}[\rho,j]
\label{eq:variational}
\end{equation}
where the optimization is over all paths $(\rho(x,t),j(x,t))$ satisfying the conditions mentioned earlier.

\begin{figure}
 \centering \includegraphics[width=0.8\textwidth]{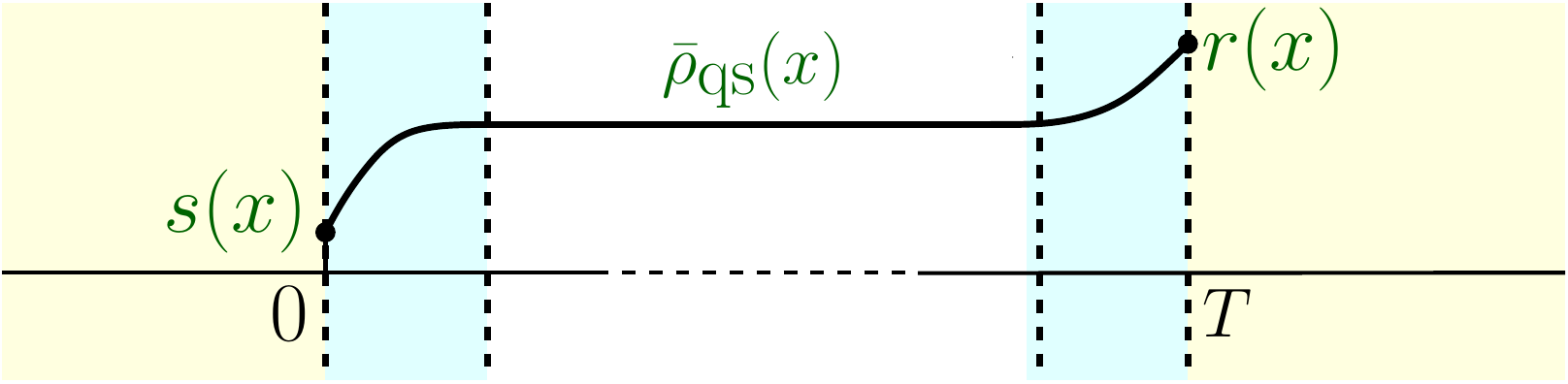}
  \caption{A schematic of optimal evolution of density for the variatonal problem \eqref{eq:variational}, where at the intermediate time the density is time independent $\bar{\rho}_\text{qs}(x)$.  \label{fig:fig4}}
\end{figure}

In a rather general class of systems \cite{Bertini2005Current}, the optimal path for \eqref{eq:variational} starts at the given density profile $\rho(x,0)=s(x)$ but soon it becomes time independent $\rho(x,t)=\bar{\rho}_\text{qs}(x)$ and remains at this density until only close to the final time $T$ where it changes to $\rho(x,T)=r(x)$. (This assumption for the time independence of the optimal profile for $t\gg 1$ and $T-t\gg 1$ is equivalent to assuming the additivity principle \cite{Bodineau2004}.) This is illustrated in the schematic in \fref{fig:fig4}. In this paper, we shall only consider situations where this scenario holds. For examples where this breaksdown see \cite{Bertini2007Current,Bodineau2007}.

\begin{remarks}~\\
\begin{enumerate}
\item The probability \eqref{eq:prob weight} does not include the contribution of reservoirs. One way to do it is to consider density profiles $\rho(x)$ which are fixed at the boundary, \textit{i.e.} $\rho(0,t)=\rho_a$ and $\rho(1,t)=\rho_b$. This is justified due to the strong coupling with the reservoirs, so that fluctuations of density at the boundary relax to the reservoir density in a time scale much faster than the hydrodynamic time scale.
\item
The formula \eqref{eq:variational} means
\begin{equation}
\chi(\kappa)=\lim_{T\rightarrow\infty}\frac{1}{T}\max_{\rho,j} S_T^{(\kappa,\alpha)}[\rho,j]
\label{eq:variational chi}
\end{equation}
which leads to the well-known result \cite{Bodineau2004,Derrida2007}
\begin{equation}
\chi(\kappa)=\max_{q}\left\{\kappa \; q- \Phi(q)\right\};\quad \Phi(q)=\min_{\bar{\rho}_\text{qs}}\int_0^1dx\frac{\left(q+D(\bar{\rho}_\text{qs})\partial_x\bar{\rho}_\text{qs}(x)\right)^2}{2\sigma(\bar{\rho}_\text{qs})} \label{eq:chi gen sol}
\end{equation}
Their solution \cite{Bodineau2004,Derrida2007} for the independent particles and for the symmetric simple exclusion process are in agreement with \eqref{eq:chi ni explicit} and \eqref{eq:chi sep expansion}.
\end{enumerate}
\end{remarks}

\subsection{Hamilton-Jacobi equation \label{sec:HJ}}
In \eqref{eq:variational}, the deviation of the optimal path from $\bar{\rho}_\text{qs}(x)$ near $t=0$ and $t=T$ (see \fref{fig:fig4}) are important and they contribute to $\psi^{(\kappa, \alpha)}_\rt$ and $\psi^{(\kappa, \alpha)}_\lt$. Here, we show how this variational formula \eqref{eq:variational} leads to a pair of Hamilton-Jacobi equations for $\psi^{(\kappa, \alpha)}_\rt$ and $\psi^{(\kappa, \alpha)}_\lt$.

We start by deriving the equation for $\psi^{(\kappa, \alpha)}_\lt$. For this, we use 
\begin{equation}
G_T^{(\kappa ,\alpha)}[r(x), s(x)]=\int\mathcal{D}[\rho]\; G_{T-t}^{(\kappa ,\alpha)}[r(x), \rho(x)]\; G_t^{(\kappa ,\alpha)}[\rho(x), s(x)]
\label{eq:time convolution}
\end{equation}
for $0< t< T$, which can be seen from the definition \eqref{eq:G def}.
We consider infinitesimal $t>0$ but large $T$, such that $T-t$ is large. This means, we can use \eqref{eq:G ldf} for $G_{T-t}^{(\kappa,\alpha)}[r(x), \rho(x)]$. On the other hand, using the Action formulation \eqref{eq:action together} we write, for an infinitesimal $t$,
\begin{equation*}
G_t^{(\kappa ,\alpha)}[\rho(x), s(x)]\sim \exp\left[t\,L\int_0^1dx\left\{\kappa\,\alpha(x) j(x)- \frac{(j(x)+D(s)\partial_x s(x))^2}{2\sigma(s)}\right\}\right]
\end{equation*}
where $\rho(x)\simeq s(x)-t\, \partial_x j(x)$. Using this in \eqref{eq:time convolution} and a saddle point analysis for large $L$ we get
\begin{align*}
\psi^{(\kappa, \alpha)}_\lt[s(x)]\simeq t\chi(\kappa)-&\max_{j(x)}\left\{-\psi^{(\kappa, \alpha)}_\lt[s(x)-t\, \partial_x j(x)]\right.\cr
&\left.+t\int_0^1dx\left(\kappa\,\alpha(x) j(x)- \frac{(j(x)+D(s)\partial_x s(x))^2}{2\sigma(s)}\right) \right\}
\end{align*}
Expanding $\psi^{(\kappa, \alpha)}_\lt[s(x)-t\, \partial_x j(x)]$ in a Taylor series up to linear order in $t$, and then using an integration by parts, we get
\begin{align}
\chi(\kappa)\simeq &\max_{j(x)}\left\{\frac{\delta \psi^{(\kappa, \alpha)}_\lt}{\delta s(1)} j(1)-\frac{\delta \psi^{(\kappa, \alpha)}_\lt}{\delta s(0)} j(0)\right.\cr
&\left.+\int_0^1dx\left[\left(\kappa\,\alpha(x)-\partial_x \frac{\delta \psi^{(\kappa, \alpha)}_\lt}{\delta s(x)}\right) j(x)- \frac{(j(x)+D(s)\partial_x s(x))^2}{2\sigma(s)}\right] \right\}
\label{eq:intermediate step opt curr}
\end{align}
For density profiles which are fixed at the boundary (see the remark 1 in \sref{sec:variational approach}), one can see from (\ref{eq:psi left no alpha together},\,\ref{eq:psi left ni together},\,\ref{eq:psi left compact formula}) that
\begin{equation}
\frac{\delta \psi^{(\kappa, \alpha)}_\lt}{\delta s(x)}=0\qquad \textrm{at $x=0$ and at $x=1$,}
\label{eq:intermediate chi}
\end{equation}
for the two systems we study in this paper. Similar conditions occured already in earlier works \cite{Bertini2002,Bertini2014}.
Using this in \eqref{eq:intermediate step opt curr} and optimizing over $j(x)$ leads to
\begin{subequations}
\begin{equation}
\int_0^1dx\left[\frac{\sigma(s)}{2}\left(\partial_x \frac{\delta \psi^{(\kappa, \alpha)}_\lt}{\delta s(x)}-\kappa\,\alpha(x)+\frac{D(s)s'(x)}{\sigma(s)}\right)^2 - \frac{(D(s) s'(x))^2}{2\sigma(s)}\right]=\chi(\kappa)
\label{eq:HJ1 canonical}
\end{equation}
where we denote $s'(x)\equiv\partial_xs(x)$.

A similar analysis (by considering small decrement around the time $T$) leads to an analogous equation for $\psi^{(\kappa, \alpha)}_\rt$ .
\begin{equation}
\int_0^1dx\left[\frac{\sigma(r)}{2}\left(\partial_x \frac{\delta \psi^{(\kappa, \alpha)}_\rt}{\delta r(x)}+\kappa\,\alpha(x)-\frac{D(r)r'(x)}{\sigma(r)}\right)^2 - \frac{(D(r) r'(x))^2}{2\sigma(r)}\right]=\chi(\kappa)
\label{eq:HJ2 canonical}
\end{equation}
\label{eq:HJ2 canonical together}
\end{subequations}
These two are the Hamilton-Jacobi equations associated to the variational problem \eqref{eq:variational}.

\subsection{Optimal path}
In \eqref{eq:intermediate step opt curr} the optimal current
\begin{equation}
j_\text{opt}(x)=-D(s)\partial_x s(x)+\sigma(s)\left(\kappa\,\alpha(x)-\partial_x \frac{\delta \psi^{(\kappa, \alpha)}_\lt}{\delta s(x)}\right)
\label{eq:Jopt}
\end{equation}
This means that the optimal path $\rho_\textrm{opt}(x,t)$ near $t=0$ follows
\begin{equation}
\partial_t \rho_\textrm{opt}=\partial_x \left\{D(\rho_\textrm{opt})\partial_x \rho_\textrm{opt}-\sigma(\rho_\textrm{opt})\left(\kappa\,\alpha(x)-\partial_x \frac{\delta \psi^{(\kappa, \alpha)}_\lt}{\delta \rho_\textrm{opt}}\right)\right\}
\label{eq:curr t 0}
\end{equation}
It is straightforward to extend the argument for $t\ge 0$ but $T-t\gg 1$ (region II of \fref{fig:fig2}) and show that the dynamics is the same.

A similar analysis in the derivation of \eqref{eq:HJ2 canonical} shows that the optimal path $\rho_\textrm{opt}(x,t)$ in region IV of \fref{fig:fig2} is described by
\begin{equation}
\partial_t \rho_\textrm{opt}=\partial_x\left\{D(\rho_\textrm{opt})\partial_x \rho_\textrm{opt}-\sigma(\rho_\textrm{opt})\left(\kappa\,\alpha(x)+\partial_x \frac{\delta \psi^{(\kappa, \alpha)}_\rt}{\delta \rho_\textrm{opt}}\right)\right\}
\label{eq:time reversed}
\end{equation}

In the quasi-stationary state \eqref{eq:time reversed} also describes the optimal path leading to a fluctuation and \eqref{eq:curr t 0} describes the optimal path of relaxation (see illustration in \fref{fig:fig3}).

\subsection{Fixed point of the dynamics}
Using \eqref{eq:HJ2 canonical together} we show in \aref{sec:H theorem} that along the optimal path \eqref{eq:curr t 0},
\begin{equation}
\frac{d}{dt}\psi_\textrm{qs}^{(\kappa)}[\rho_\textrm{opt}(x,t)]=-\int_0^1dx\,\frac{\sigma(\rho_\textrm{opt}(x,t))}{2}\left( \partial_x \frac{\delta \psi_\textrm{qs}^{(\kappa)}[\rho_\textrm{opt}(x,t)]}{\delta \rho_\textrm{opt}(x,t)}\right)^2 \label{eq:H theorem}
\end{equation}
with $\psi_\textrm{qs}^{(\kappa)}$ given in \eqref{eq:psi mid reln 2}. Since $\sigma(\rho)$ is positive, this means $\frac{d}{dt}\psi_\textrm{qs}^{(\kappa)}[\rho_\textrm{opt}(x,t)]=0$ if and only if $\partial_x\frac{\delta \psi_\textrm{qs}^{(\kappa)}[\rho_\textrm{opt}(x,t)]}{\delta \rho_\textrm{opt}(x,t)}=0$. (The case $\kappa=0$ has been discussed earlier in \cite{Bertini2002}.)

The examples we consider here have a unique quasi-stationary density $\bar{\rho}_\textrm{qs}$, where $\frac{\delta \psi_\textrm{qs}^{(\kappa)}}{\delta \bar \rho_\text{qs}(x)}=0$. Then \eqref{eq:H theorem} implies that $\bar{\rho}_\textrm{qs}$ is an attractive fixed point of \eqref{eq:curr t 0} (see \fref{fig:fig3}).

For $\bar{\rho}_\textrm{qs}$ the optimal current \eqref{eq:Jopt} is $j_\text{opt}=\chi'(\kappa)$, which can be seen from \eqref{eq:chi gen sol}. Then, we get
\begin{align*}
-D(\bar{\rho}_\textrm{qs})\partial_x \bar{\rho}_\textrm{qs}+\sigma(\bar{\rho}_\textrm{qs})\left(\kappa\,\alpha(x)-\partial_x \frac{\delta \psi^{(\kappa, \alpha)}_\lt}{\delta \bar{\rho}_\textrm{qs}(x)}\right)&=\chi'(\kappa)
\end{align*}
which leads to
\begin{subequations}
\begin{equation}
\partial_x \frac{\delta \psi^{(\kappa, \alpha)}_\lt}{\delta \bar{\rho}_\textrm{qs}(x)}=\kappa\,\alpha(x)-\frac{\chi'(\kappa)+D(\bar{\rho}_\textrm{qs})\partial_x \bar{\rho}_\textrm{qs}(x)}{\sigma(\bar{\rho}_\textrm{qs})}
\end{equation}
A similar calculation for \eqref{eq:time reversed} lead to
\begin{equation}
\partial_x \frac{\delta \psi^{(\kappa, \alpha)}_\rt}{\delta \bar{\rho}_\textrm{qs}(x)} =-\kappa\,\alpha(x)+\frac{\chi'(\kappa)+D(\bar{\rho}_\textrm{qs})\partial_x \bar{\rho}_\textrm{qs}(x)}{\sigma(\bar{\rho}_\textrm{qs})}
\end{equation}
\label{eq:proposition 1}
\end{subequations}

These give conditions for the solution of \eqref{eq:HJ2 canonical together}. It is well-known \cite{Goldstein2000,Bertini2002} that, there are multiple solutions of a Hamilton-Jacobi equation. For the two examples studied in this work, the relevant solution of \eqref{eq:HJ2 canonical together} follows the boundary condition \eqref{eq:proposition 1} and (see \eqref{eq:intermediate chi})
\begin{align}
\frac{\delta \psi^{(\kappa, \alpha)}_\lt}{\delta \rho(x)}=0\quad \textrm{and} \quad \frac{\delta \psi^{(\kappa, \alpha)}_\rt}{\delta \rho(x)}=0 \qquad \textrm{at $x=0$ and $x=1$.} \label{eq:psi spatial boundary}
\end{align}

\subsection{Conditioned stochastic dynamics \label{sec:conditioned macro}}
In \sref{sec:conditioned dynamics} we have shown using a microscopic analysis that, in the quasi-stationary state, conditioned dynamics is given by a fluctuating hydrodynamics equation \eqref{eq:fhd fluc together}. Here, we give a derivation using the macroscopic approach.

In the quasi-stationary state, if $\mathcal{P}[\rho,j\vert \rho_i]$ is the probability of a path $\left\{\rho(x,t),j(x,t)\right\}$ in a time window $[t_i,t_f]$ (for $1\ll t_i<t_f$ and $1\ll T-t_f$) given an initial density $\rho(x,t_i)=\rho_i(x)$, then using \eqref{eq:G def} and \eqref{eq:action} one can write
\begin{equation*}
\mathcal{P}[\rho,j\vert \rho_i]=\frac{G_{T-t_f}^{(\kappa,\alpha)}[\rho_T,\rho_f]\; e^{L S_{t_f-t_i}^{(\kappa,\alpha)}}}{G_{T-t_i}^{(\kappa,\alpha)}[\rho_T,\rho_i]}
\end{equation*}
where we denote $\rho(x,t_f)=\rho_f(x)$ and $\rho(x,T)=\rho_T(x)$.
Then, for large $L$, using \eqref{eq:psit general} and \eqref{eq:G ldf} we get
\begin{equation}
\mathcal{P}[\rho,j\vert \rho_i]\sim e^{-L (t_f-t_i)\chi(\kappa)+L \psi_\lt^{(\kappa,\alpha)}[\rho_i]-L \psi_\lt^{(\kappa,\alpha)}[\rho_f]+ L S_{t_f-t_i}^{(\kappa,\alpha)}[\rho,j]}
\label{eq:P cond prelim}
\end{equation}
This expression can be simplified by using 
\begin{align*}
\psi_\lt^{(\kappa,\alpha)}[\rho_f]-\psi_\lt^{(\kappa,\alpha)}[\rho_i]=&\int_0^1 dx \int_{t_i}^{t_f}dt\, \partial_t\rho\,\frac{\delta \psi_\lt^{(\kappa,\alpha)}[\rho]}{\delta \rho(x,t)}\cr
=& \int_0^1 dx \int_{t_i}^{t_f}dt\, j(x,t)\,\partial_x\left(\frac{\delta \psi_\lt^{(\kappa,\alpha)}[\rho]}{\delta \rho(x,t)}\right)
\end{align*}
where the last equality is obtained by using $\partial_t\rho=-\partial_x j$, integration by parts, and the boundary condition \eqref{eq:psi spatial boundary}. In addition, we use \eqref{eq:HJ1 canonical} to write
\begin{subequations}
\begin{equation*}
(t_f-t_i)\chi(\kappa)=\int_0^1dx\int_{t_i}^{t_f}dt\left[\frac{\sigma(\rho)}{2}\left(\partial_x \frac{\delta \psi^{(\kappa, \alpha)}_\lt}{\delta \rho(x,t)}-\kappa\,\alpha(x)+\frac{D(\rho)\partial_x\rho(x,t)}{\sigma(\rho)}\right)^2 - \frac{(D(\rho)\partial_x \rho(x))^2}{2\sigma(\rho)}\right]
\end{equation*}
Using the above two results in \eqref{eq:P cond prelim} and following a simple algebra we get
\begin{equation}
\mathcal{P}[\rho,j\vert \rho_i]\sim e^{L \,\widehat{S}[\rho,j]}
\label{eq:P cond action}
\end{equation}
with the Action
\begin{equation}
\widehat{S}[\rho,j]=-\int_{t_i}^{t_f}dt\int_0^1dx\frac{\left\{j(x,t)+D(\rho)\partial_x\rho(x,t)+\sigma(\rho)\left(\partial_x \frac{\delta \psi^{(\kappa, \alpha)}_\lt}{\delta \rho(x,t)}-\kappa\,\alpha(x) \right)\right\}^2}{2\sigma(\rho)}
\label{eq:conditioned Action}
\end{equation}
\label{eq:conditioned Action together}
\end{subequations}

Comparing with \eqref{eq:prob weight} one can clearly see that the conditioned dynamics in the quasi-stationary state is given by a fluctuating hydrodynamics equation $\partial_t\rho(x,t)=-\partial_x j(x,t)$ with $j(x,t)$ in \eqref{eq:fhd relax}. This describes, how a spontaneous fluctuation relaxes in the quasi-stationary state.

On the other hand, \eqref{eq:fhd fluc} shows the path leading to a fluctuation. This is given by a time reversal of \eqref{eq:fhd relax}, which can be constructed (for example see eq. 2.15 of \cite{Bertini2002}) by using that $P_\textrm{qs}^{(\kappa,\alpha)}[\rho]$ is the steady state of \eqref{eq:fhd relax}. This gives a fluctuating hydrodynamics equation $\partial_t\rho(x,t)=-\partial_x \left\{j(x,t)+\sigma(\rho)\partial_x\frac{\delta \psi_\textrm{qs}^{(\kappa)}}{\delta \rho(x,t)}\right\}$ with the $j(x,t)$ in \eqref{eq:fhd relax}. Then, using \eqref{eq:psi mid reln 2} one gets \eqref{eq:fhd fluc}.

\section{Solution in specific examples \label{sec:examples macro}}
Here, we show how to check that the results for $\psi_\lt^{(\kappa,\alpha)}$ and $\psi_\rt^{(\kappa,\alpha)}$ derived in \sref{sec:ni micro analysis} and \sref{sec:sep micro analysis} using a microscopic analysis, are indeed solution of the Hamilton-Jacobi equations \eqref{eq:HJ2 canonical together}.

\subsection{Independent particles \label{sec:ni macro}}
In this case, using \eqref{eq:psi left no alpha} and \eqref{eq:chi ni explicit} in \eqref{eq:HJ1 canonical} we get
\begin{equation}
\int_0^1dx\left[\rho(x)\left(\partial_x \frac{\delta V^{(\kappa)}_\lt}{\delta \rho(x)}\right)^2+\rho'(x)\left(\partial_x \frac{\delta V^{(\kappa)}_\lt}{\delta \rho(x)}\right)\right]=\rho_a\left(e^{\kappa}-1 \right)+\rho_b\left(e^{-\kappa}-1 \right)
\label{eq:HJ1 canonical no alpha app}
\end{equation}
Expression \eqref{eq:psi left ni} for $V^{(\kappa)}_\lt$ gives
\begin{align*}
\partial_x \frac{\delta V^{(\kappa)}_\lt}{\delta \rho(x)}= \frac{(e^{-\kappa}-1)}{b(x)}\qquad\textrm{and}\qquad 
\left(\partial_x \frac{\delta V^{(\kappa)}_\lt}{\delta \rho(x)}\right)^2=(e^{-\kappa}-1)\left(\frac{1}{b(x)}\right)'
\end{align*}
With this, the left hand side of \eqref{eq:HJ1 canonical no alpha app} becomes
\begin{align*}
(e^{-\kappa}-1)\int_0^1dx\left[\rho(x)\left(\frac{1}{b(x)}\right)'+\frac{\rho'(x)}{b(x)}\right]=&(e^{-\kappa}-1)\int_0^1dx\left(\frac{\rho(x)}{b(x)}\right)'\cr=&(e^{-\kappa}-1)\left(\frac{\rho(1)}{b(1)}-\frac{\rho(0)}{b(0)}\right)
\end{align*}
From the boundary condition $\rho(0)=\rho_a$, $\rho(1)=\rho_b$, and using \eqref{eq:ell 0} we see that the above expression agrees with the right hand side of \eqref{eq:HJ1 canonical no alpha app}.

Moreover, one can check that the solution (\ref{eq:psi left no alpha},\,\ref{eq:psi left ni}) is consistent with the boundary condition (\ref{eq:proposition 1},\,\ref{eq:psi spatial boundary}).

A similar calculation shows that $\psi_\rt^{(\kappa)}$ in (\ref{eq:psi right no alpha},\,\ref{eq:psi right ni}) is a solution of \eqref{eq:HJ2 canonical} with the boundary condition (\ref{eq:proposition 1},\,\ref{eq:psi spatial boundary}).

\subsection{Symmetric simple exclusion process \label{sec:sep macro}}
In this case, our solution for $\psi_\lt^{(\kappa,\alpha)}$ and $\psi_\rt^{(\kappa,\alpha)}$ in (\ref{eq:psi left no alpha together},\,\ref{eq:psi left compact formula},\,\ref{eq:psif compact}) are for small density. We have explicitly verified, up to the second order in density, that these are solutions of the Hamilton-Jacobi equations \eqref{eq:HJ2 canonical together} and they satisfy the boundary condition (\ref{eq:proposition 1},\,\ref{eq:psi spatial boundary}). The analysis is similar to that of the independent particles in \sref{sec:ni macro}. In fact, at the leading order in density, they are identical.

\section{Summary}
In the present work we have tried to determine the probability of the density \eqref{eq:cond prob micro} in a diffusive many-particle system conditioned on the time-integrated current \eqref{eq:QT} for large $T$. This is a generalization to extended systems of earlier works on conditioned stochastic processes \cite{Sadhu20182,Jack2015,Chetrite2013,Chetrite2015,TOUCHETTE2017,Lecomte20072,Garrahan2009}. We mostly worked with the canonical ensemble where dynamics is weighted by the current \eqref{eq:QT}. However, the equivalence of ensembles allows to make predictions for the conditioned process (see \sref{sec:examples microscopic}). We give explicit results for a system of independent particles (\sref{sec:ni micro analysis}) and the symmetric simple exclusion process (\sref{sec:sep micro analysis}).

In the hydrodynamic limit, the conditioned probability of the density $\rho(x)$ is characterized by the large deviation function $\psi_t^{(\kappa,\alpha)}[\rho]$ in \eqref{eq:psit general}. For the two systems considered in this paper, we have calculated $\psi_t^{(\kappa,\alpha)}[\rho]$ at three different times of the evolution, namely, at $t=0$, at $t=T$, and in the quasi-stationary regime. These are, in general, related by (see \eqref{eq:psi 0 canonical together})
\begin{equation}
\psi_0^{(\kappa,\alpha)}[\rho]+\psi_T^{(\kappa,\alpha)}[\rho]=\psi_\text{qs}^{(\kappa,\alpha)}[\rho]+\mathcal{F}[\rho]\label{eq:all psi rel}
\end{equation}
with $\mathcal{F}$ defined in \eqref{eq:ldf individual}.

In the second half of the paper, we used a macroscopic approach, where $\psi_t^{(\kappa,\alpha)}$ is expressed (see \eqref{eq:psi 0 canonical together}) in terms of $\psi_\lt^{(\kappa,\alpha)}$ and $\psi_\rt^{(\kappa,\alpha)}$, which are solutions of a pair of Hamilton-Jacobi equations \eqref{eq:HJ2 canonical together}. These solutions also act as the potential for an additional field that drives the conditioned process (see \eqref{eq:fhd fluc together} and \sref{sec:conditioned macro}). Using this macroscopic approach we verified the microscopic results for the two specific examples (see \sref{sec:examples macro}).

The macroscopic formulation is expected to work for a wide class of diffusive systems where the microscopic details enter in the two functions $D(\rho)$ and $\sigma(\rho)$ in \eqref{eq:fhd together}. This is in the spirit of the macroscopic fluctuation theory \cite{Bertini2014}. It would be interesting find more examples for explicit solutions of $\psi_\lt^{(\kappa,\alpha)}$ and $\psi_\rt^{(\kappa,\alpha)}$ using both the microscopic and the macroscopic approach. For systems on a ring, where the unconditioned state is in equilibrium, the calculation may be simpler. However, due to the periodic boundary condition, the optimal profile in the quasi-stationary state could become time dependent (see \cite{Bertini2007Current,Bodineau2007,Nicolas2018}).

In the unconditioned case ($\kappa=0$), the spatial correlations of density follow simple differential equations \cite{Sadhu2016,Bertini2007Correlation}, whose solutions can be formally expressed in terms of a Green's function. It would be interesting to see if there are similar equations for the conditioned case, especially in the quasi-stationary state.

\begin{acknowledgements}
We thank T. Bodineau for his useful comments related to \eqref{eq:psi mid reln 2} and \eqref{eq:all psi rel}.
\end{acknowledgements}

\appendix

\section{Tilted matrix for the symmetric simple exclusion process \label{sec:eigen equation sep}}
Here, we explicitly write the eigenvalue equation for the tilted Matrix in a symmetric simple exclusion process of arbitrary length $L$. Our results are for the case $\lambda_0=1$ and $\lambda_i=0$ for $i\ge 1$. We use the representation \eqref{eq:eigenvector sep representation} for the eigenvectors with a normalization such that the component of both right and left eigenvectors for the empty configuration is $1$. We write the eigenvalue equation up to the two particle sector.

For the right eigenvector, defining $\mathcal{R}^{(\kappa)}(i,j)=0$ for $i=j$, we get the following set of coupled equations.
\begin{itemize}
\item Empty-particle sector.
\begin{equation*}
\mu+\rho_a+\rho_b=e^{-\kappa}(1-\rho_a)\mathcal{R}^{(\kappa)}(1)+(1-\rho_b)\mathcal{R}^{(\kappa)}(L)
\end{equation*}
\item
Single-particle sector.
\begin{itemize}
\item[$\circ$] For $1<i<L$,
\begin{align*}
(\mu+\rho_a+\rho_b)\mathcal{R}^{(\kappa)}(i)-&\left[\mathcal{R}^{(\kappa)}(i-1)-2\mathcal{R}^{(\kappa)}(i)+\mathcal{R}^{(\kappa)}(i+1)\right]\cr
=&e^{-\kappa}(1-\rho_a)\mathcal{R}^{(\kappa)}(1,i)+(1-\rho_b)\mathcal{R}^{(\kappa)}(i,L)
\end{align*}

\item[$\circ$] For $i=1$,
\begin{align*}
(\mu+\rho_a+\rho_b)\mathcal{R}^{(\kappa)}(1)-&\left[2\rho_a\mathcal{R}^{(\kappa)}(1)-2\mathcal{R}^{(\kappa)}(1)+\mathcal{R}^{(\kappa)}(2)\right]\cr
=&e^{\kappa}\rho_a+(1-\rho_b)\mathcal{R}^{(\kappa)}(1,L)
\end{align*}

\item[$\circ$] For $i=L$,
\begin{align*}
(\mu+\rho_a+\rho_b)\mathcal{R}^{(\kappa)}(L)-&\left[\mathcal{R}^{(\kappa)}(L-1)-2\mathcal{R}^{(\kappa)}(L)+2\rho_b\mathcal{R}^{(\kappa)}(L)\right]\cr=&e^{-\kappa}(1-\rho_a)\mathcal{R}^{(\kappa)}(1,L)+\rho_b
\end{align*}
\end{itemize}
\item 
Two-particle sector.
\begin{itemize}
\item[$\circ$] For $1<i<j<L$,
\begin{align*}
(\mu+\rho_a+\rho_b)&\mathcal{R}^{(\kappa)}(i,j)-\bigg[\mathcal{R}^{(\kappa)}(i-1,j)+\mathcal{R}^{(\kappa)}(i,j-1)-4\mathcal{R}^{(\kappa)}(i,j)\cr
&\qquad +\mathcal{R}^{(\kappa)}(i+1,j)+\mathcal{R}^{(\kappa)}(i,j+1)\bigg]
-2\mathcal{R}^{(\kappa)}(i,j)\delta_{i+1,j} \cr &=e^{-\kappa}(1-\rho_a)\mathcal{R}^{(\kappa)}(1,i,j)+(1-\rho_b)\mathcal{R}^{(\kappa)}(i,j,L)
\end{align*}

\item[$\circ$]For $1=i<j<L$,
\begin{align*}
(\mu+\rho_a&+\rho_b)\mathcal{R}^{(\kappa)}(1,j)-\bigg[2\rho_a\mathcal{R}^{(\kappa)}(1,j)+\mathcal{R}^{(\kappa)}(1,j-1)-4\mathcal{R}^{(\kappa)}(1,j)+\mathcal{R}^{(\kappa)}(2,j)\cr
&+\mathcal{R}^{(\kappa)}(1,j+1)\bigg]-2\mathcal{R}^{(\kappa)}(1,j)\delta_{2,j} =e^{\kappa}\rho_a\mathcal{R}^{(\kappa)}(j)+(1-\rho_b)\mathcal{R}^{(\kappa)}(1,j,L)
\end{align*}

\item[$\circ$]For $1<i<j=L$,
\begin{align*}
(\mu+\rho_a&+\rho_b)\mathcal{R}^{(\kappa)}(i,L)-\bigg[\mathcal{R}^{(\kappa)}(i-1,L)+\mathcal{R}^{(\kappa)}(i,L-1)-4\mathcal{R}^{(\kappa)}(i,L)+\mathcal{R}^{(\kappa)}(i+1,L)\cr
&+2\rho_b\mathcal{R}^{(\kappa)}(i,L)\bigg]-2\mathcal{R}^{(\kappa)}(i,L) \delta_{i,L-1}=e^{-\kappa}(1-\rho_a)\mathcal{R}^{(\kappa)}(1,i,L)+\rho_b\mathcal{R}^{(\kappa)}(i)
\end{align*}
\end{itemize}
\end{itemize}

Similarly, for the left eigenvector, we define $\mathcal{L}^{(\kappa)}(i,j)=0$ for $i=j$, and get the following set of equations.
\begin{itemize}
\item Empty particle sector.
\begin{equation*}
\mu+\rho_a+\rho_b=e^{\kappa}\rho_a\mathcal{L}^{(\kappa)}(1)+\rho_b\mathcal{L}^{(\kappa)}(L)
\end{equation*}
\item
Single particle sector.
\begin{itemize}
\item[$\circ$] For $1<i<L$,
\begin{align*}
(\mu+\rho_a+\rho_b)\mathcal{L}^{(\kappa)}(i)-&\left[\mathcal{L}^{(\kappa)}(i-1)-2\mathcal{L}^{(\kappa)}(i)+\mathcal{L}^{(\kappa)}(i+1)\right]\cr
=&e^{\kappa}\rho_a\mathcal{L}^{(\kappa)}(1,i)+\rho_b\mathcal{L}^{(\kappa)}(i,L)
\end{align*}

\item[$\circ$] For $i=1$,
\begin{align*}
(\mu+\rho_a+\rho_b)\mathcal{L}^{(\kappa)}(1)-&\left[2\rho_a\mathcal{L}^{(\kappa)}(1)-2\mathcal{L}^{(\kappa)}(1)+\mathcal{L}^{(\kappa)}(2)\right]\cr
=&e^{-\kappa}(1-\rho_a)+\rho_b\mathcal{L}^{(\kappa)}(1,L)
\end{align*}

\item[$\circ$] For $i=L$,
\begin{align*}
(\mu+\rho_a+\rho_b)\mathcal{L}^{(\kappa)}(L)-&\left[\mathcal{L}^{(\kappa)}(L-1)-2\mathcal{L}^{(\kappa)}(L)+2\rho_b\mathcal{L}^{(\kappa)}(L)\right]\cr\quad =&e^{\kappa}\rho_a\mathcal{L}^{(\kappa)}(1,L)+(1-\rho_b)
\end{align*}
\end{itemize}
\item 
Two particle sector.
\begin{itemize}
\item[$\circ$] For $1<i<j<L$,
\begin{align*}
(\mu+\rho_a+\rho_b)&\mathcal{L}^{(\kappa)}(i,j)-\bigg[\mathcal{L}^{(\kappa)}(i-1,j)+\mathcal{L}^{(\kappa)}(i,j-1)-4\mathcal{L}^{(\kappa)}(i,j)\cr
&\qquad +\mathcal{L}^{(\kappa)}(i+1,j)+\mathcal{L}^{(\kappa)}(i,j+1)\bigg]-2\mathcal{L}^{(\kappa)}(i,j)\delta_{i+1,j}\cr & \qquad\qquad\quad =e^{\kappa}\rho_a\mathcal{L}^{(\kappa)}(1,i,j)+\rho_b\mathcal{L}^{(\kappa)}(i,j,L)
\end{align*}

\item[$\circ$]For $1=i<j<L$,
\begin{align*}
(\mu+\rho_a&+\rho_b)\mathcal{L}^{(\kappa)}(1,j)-\bigg[2\rho_a\mathcal{L}^{(\kappa)}(1,j)+\mathcal{L}^{(\kappa)}(1,j-1)-4\mathcal{L}^{(\kappa)}(1,j)+\mathcal{L}^{(\kappa)}(2,j)\cr
&+\mathcal{L}^{(\kappa)}(1,j+1)\bigg]-2\mathcal{L}^{(\kappa)}(1,j)\delta_{2,j} =e^{-\kappa}(1-\rho_a)\mathcal{L}^{(\kappa)}(j)+\rho_b\mathcal{L}^{(\kappa)}(1,j,L)
\end{align*}

\item[$\circ$]For $1<i<j=L$,
\begin{align*}
(\mu+\rho_a&+\rho_b)\mathcal{L}^{(\kappa)}(i,L)-\bigg[\mathcal{L}^{(\kappa)}(i-1,L)+\mathcal{L}^{(\kappa)}(i,L-1)-4\mathcal{L}^{(\kappa)}(i,L)+\mathcal{L}^{(\kappa)}(i+1,L)\cr
&+2\rho_b\mathcal{L}^{(\kappa)}(i,L)\bigg]-2\mathcal{L}^{(\kappa)}(i,L)\delta_{i,L-1}
=e^{\kappa}\rho_a\mathcal{L}^{(\kappa)}(1,i,L)+(1-\rho_b)\mathcal{L}^{(\kappa)}(i)
\end{align*}
\end{itemize}
\end{itemize}

It is easy to see a pattern in the equations and using this we can write the equations for an arbitrary particle sector. It is then possible to systematically solve the equations order by order using the perturbation expansion in small density as given in \sref{sec:perturbation expansion sep}.

\section{Cumulants for the symmetric simple exclusion process \label{app:cumulants}}
We write the cumulants of density using a set of parameters 
\begin{equation}
z=e^{\kappa},\qquad h=e^{\kappa}-1, \qquad s=\rho_a h, \qquad \textrm{and} \qquad p=1-\frac{\rho_b}{\rho_a\,z}
\end{equation}
which was used earlier \cite{Roche2004}. In terms of these, the cumulants in \eqref{eq:rho c} have the following expression.

\begin{itemize}
\item
At $t=0$,
\begin{align*}
\bar{\rho}(x)&= \frac{\rho_a}{z} \bigg\{(1+xh)(1+xh-(pz) x) +\frac{s}{6z}(1-x)\bigg[6+2xh\big(7-z+xh(6-z+2xh)\big)\cr
&\qquad -(pz)x\big(10+2z+xh(16-2z+7xh) \big) +(pz)^2 x\big(2+x(5+3xh) \big)\bigg]+\mathcal{O}(s^2)\bigg\},\cr
c(x,y)&= -\frac{\rho_a^2}{z^2} x(1-y)\bigg\{2h^2(1+xh)(1+yh) -h(pz)\big( 2+3h(x+y)+4h^2xy \big)\cr
& \qquad \qquad \qquad \qquad \qquad +(pz)^2\big(1+h(x+y)+2h^2xy \big)+\mathcal{O}(s)\bigg\}\qquad \text{for $x\le y$.}
\end{align*}
\item
In the quasi-stationary state,
\begin{align*}
\bar{\rho}(x)&\equiv\bar{\rho}_\text{qs}(x) =\rho_a \bigg\{(1-px)(1+xh)\cr
&\qquad \qquad -sx(1-x)\bigg[ h +\frac{p}{3}(7-z+2xh)-\frac{p^2}{3}(1+x)(2+xh)\bigg]+\mathcal{O}(s^2)\bigg\}, \cr
c(x,y)&= -\rho_a^2 x(1-y)\bigg\{p^2(1+hx)(1+hy)+h^2(1-px)(1-py)+\mathcal{O}(s)\bigg\} \qquad \text{for $x\le y$.}
\end{align*}
\item
at $t=T$,
\begin{align*}
\bar{\rho}(x)&=\rho_a \, z \bigg\{1-px+s\frac{(1-x)}{6}\left[-6+p^2x(2-x)\right]+\mathcal{O}(s^2)\bigg\}, \cr
c(x,y)&= -\rho_a^2 z^2x(1-y) \bigg\{p^2+\mathcal{O}(s)\bigg\}\qquad \text{for $x\le y$.}
\end{align*}
\end{itemize}

\section{Small density expansion of $V_\lt^{(\kappa)}$ \label{sec:mlf}}
Here, we show that the expression of $V_\lt^{(\kappa)}$ in \eqref{eq:psi left compact formula} agrees with $\psi_\textrm{qs}^{(\kappa)}-\psi_T^{(\kappa)}$, up to an additive constant, as expected from \eqref{eq:psi mid reln 2}. To see this, we use \eqref{eq:F sep series} and we get
\begin{align}
\psi_\textrm{qs}^{(\kappa)}[\rho]-\psi_T^{(\kappa)}[\rho]= &\int_0^{1}dx \log\frac{1-\bar{\rho}_T(x)}{1-\bar{\rho}_\textrm{qs}(x)}+\frac{1}{2}\int_0^{1}dx\int_0^1 dy \left\{c_T(x,y)-c_\textrm{qs}(x,y) \right\}\cr
&+\int_0^{1}dx \, \rho(x)\log\Bigg[\frac{\bar{\rho}_T(x)}{1-\bar{\rho}_T(x)}\frac{1-\bar{\rho}_\textrm{qs}(x)}{\bar{\rho}_\textrm{qs}(x)}\Bigg]\cr
&-\int_0^{1}dx\, \rho(x)\int_0^1 dy \Bigg\{\frac{c_T(x,y)}{\bar{\rho}_T(x)}-\frac{c_\textrm{qs}(x,y)}{\bar{\rho}_\textrm{qs}(x)} \Bigg\}\cr
& -\frac{1}{2}\int_0^{1}dx\int_0^1 dy\,  \rho(x)\rho(y)\Bigg\{\frac{c_\textrm{qs}(x,y)}{\bar{\rho}_\textrm{qs}(x)\bar{\rho}_\textrm{qs}(y)}- \frac{c_T(x,y)}{\bar{\rho}_T(x)\bar{\rho}_T(y)}\Bigg\}  +\cdots \label{eq:psi qs minus T}
\end{align}
where the subscripts $T$ and $\text{qs}$ refer to the cumulants \eqref{eq:rho c} at time $t=T$ and in the quasi-stationary state. (In the third line of the above equation we used that $c_T(x,y)$ and $c_\textrm{qs}(x,y)$ are symmetric under exchange of $x$ and $y$.) The terms in the first line are constant and therefore ignored. For the rest of the terms we use
\begin{align*}
c_\textrm{qs}(x,y)\simeq &\;\ell^{(\kappa)}(x)\ell^{(\kappa)}(y)c_T(x,y)+\bar{\rho}_T(x)\bar{\rho}_T(y)g_\lt^{(\kappa)}(x,y) \cr
\bar{\rho}_\textrm{qs}(x)\simeq &\; \ell^{(\kappa)}(x)\bar{\rho}_T(x)\Big[1-\ell^{(\kappa)}(x)\bar{\rho}_T(x)\Big]+\ell^{(\kappa)}(x)\bar{\rho}_T(x)^2\cr
&-\ell^{(\kappa)}(x)\int_0^1dy\,\left(1-\ell^{(\kappa)}(y) \right)c_T(x,y)+\bar{\rho}_T(x)\int_0^1dy\,\bar{\rho}_T(y)g_\lt^{(\kappa)}(x,y)
\end{align*}
up to the second order in $\rho_a$ and $\rho_b$, which can be seen from \eqref{eq:cumulants qs} and \eqref{eq:cumulants T}.

Then, it is straightforward to see that, up to the second order in $\rho_a$ and $\rho_b$,
\begin{equation*}
\rho(x)\rho(y)\left\{\frac{c_\textrm{qs}(x,y)}{\bar{\rho}_\textrm{qs}(x)\bar{\rho}_\textrm{qs}(y)}- \frac{c_T(x,y)}{\bar{\rho}_T(x)\bar{\rho}_T(y)}\right\}\simeq\; \rho(x)\rho(y)\;\frac{g_\lt^{(\kappa)}(x,y)}{\ell^{(\kappa)}(x)\ell^{(\kappa)}(y)}
\end{equation*}
and
\begin{align*}
\log\Bigg[\frac{\bar{\rho}_T(x)}{1-\bar{\rho}_T(x)}\frac{1-\bar{\rho}_\textrm{qs}(x)}{\bar{\rho}_\textrm{qs}(x)}\Bigg]-\int_0^1 dy \Bigg\{\frac{c_T(x,y)}{\bar{\rho}_T(x)}-\frac{c_\textrm{qs}(x,y)}{\bar{\rho}_\textrm{qs}(x)} \Bigg\}\simeq\log\frac{1}{\ell^{(\kappa)}(x)}
\end{align*}
This shows that the expression in \eqref{eq:psi qs minus T} agrees with $V_\lt^{(\kappa)}$ in \eqref{eq:psi left compact formula}, up to an additive constant.

\section{A H-theorem \label{sec:H theorem}}
To derive \eqref{eq:H theorem} we write, along the optimal path \eqref{eq:curr t 0},
\begin{equation*}
\frac{d}{dt}\psi_\text{qs}^{(\kappa,\alpha)} =\int_0^1dx\,\frac{\delta\psi_\text{qs}^{(\kappa,\alpha)}}{\delta \rho_\textrm{opt}}\partial_t\rho_\textrm{opt}
\end{equation*}
Then, using \eqref{eq:curr t 0} and an integration by parts we get
\begin{align*}
\frac{d}{dt}\psi_\text{qs}^{(\kappa,\alpha)}&=-\int_0^1dx\,\left\{\partial_x\frac{\delta\psi_\text{qs}^{(\kappa,\alpha)}}{\delta \rho_\textrm{opt}} \right\}\left\{D(\rho_\textrm{opt})\partial_x\rho_\textrm{opt}+\sigma(\rho_\textrm{opt})\left(\partial_x\frac{\delta\psi_\lt^{(\kappa,\alpha)}}{\delta \rho_\textrm{opt}}-\kappa\alpha(x) \right) \right\}
\end{align*}
where we use $\frac{\delta\psi_\text{qs}^{(\kappa,\alpha)}}{\delta \rho_\textrm{opt}(x,t)}=0$ at the boundary $x=0$ and $1$ (due to \eqref{eq:psi mid reln 2} and \eqref{eq:psi spatial boundary}).
To simplify the expression, we use
\begin{equation*}
\int_0^1dx\,\left\{\partial_x\frac{\delta\psi_\text{qs}^{(\kappa,\alpha)}}{\delta \rho_\textrm{opt}} \right\}\left\{D(\rho_\textrm{opt})\partial_x\rho_\textrm{opt}+\frac{\sigma(\rho_\textrm{opt})}{2}\left(\partial_x\frac{\delta\psi_\lt^{(\kappa,\alpha)}}{\delta \rho_\textrm{opt}}-\partial_x\frac{\delta\psi_\rt^{(\kappa,\alpha)}}{\delta \rho_\textrm{opt}}-2\kappa \alpha(x)\right)\right\}=0
\end{equation*}
which is obtained by subtracting \eqref{eq:HJ2 canonical} from \eqref{eq:HJ1 canonical} and then using \eqref{eq:psi mid reln 2}.
From the above two equations we get \eqref{eq:H theorem}.

\bibliographystyle{iopart-num}
\bibliography{reference,reference_old}

\end{document}